\documentclass[11pt,a4paper]{article}
\pdfoutput=1

\usepackage{bm}
\usepackage{amsmath,amssymb}
\usepackage{cite}

\usepackage{tikz}
\usepackage{graphicx}
\usepackage{color}
\usepackage{upgreek}

\usepackage{hyperref} 
\hypersetup{colorlinks=true, citecolor=blue, filecolor=black, linkcolor=blue, urlcolor=blue, pdfpagemode=UseNone}

\numberwithin{figure}{section}
\numberwithin{equation}{section}

\newcommand{\be}{\begin{equation}}
\newcommand{\ee}{\end{equation}}
\newcommand{\bea}{\begin{eqnarray}}
\newcommand{\eea}{\end{eqnarray}}
\def\beal#1\eeal{\begin{align}#1\end{align}}   
\def\besp#1\eesp{\begin{multline}#1\end{multline}} 

\newcommand\ie{\textit{i.e.}\ }
\newcommand\eg{\textit{e.g.}\ }
\newcommand\cf{\textit{cf.}\ }

\newcommand{\etc}{{\it etc.}\ }

\newcommand{\nn}{\nonumber}

\makeatletter
\providecommand*{\shuffle}{%
  \mathbin{\mathpalette\shuffle@{}}%
}
\newcommand*{\shuffle@}[2]{%
  \sbox0{$#1\vcenter{}$}%
  \kern .15\ht0 
  \rlap{\vrule height .25\ht0 depth 0pt width 2.5\ht0}%
  \raise.1\ht0\hbox to 2.5\ht0{%
    \vrule height 1.75\ht0 depth -.1\ht0 width .17\ht0 %
    \hfill
    \vrule height 1.75\ht0 depth -.1\ht0 width .17\ht0 %
    \hfill
    \vrule height 1.75\ht0 depth -.1\ht0 width .17\ht0 %
  }%
  \kern .15\ht0 
}
\makeatother

\newcommand{\K}{\mathcal{K}}
\newcommand{\rK}{\K}
\newcommand{\lK}{\overleftarrow{\K}}
\newcommand{\lrK}{\overleftrightarrow{\K}}

\begin{document}

\begin{titlepage}

\begin{center}
{\huge \bf Properties of a proposed background independent exact renormalization group}

\end{center}
\vskip1cm


\begin{center}
{\bf Vlad-Mihai Mandric and Tim R. Morris}
\end{center}

\begin{center}
{\it STAG Research Centre \& Department of Physics and Astronomy,\\  University of Southampton,
Highfield, Southampton, SO17 1BJ, U.K.}\\
\vspace*{0.3cm}
{\tt  V.M.Mandric@soton.ac.uk, T.R.Morris@soton.ac.uk}
\end{center}

\begin{abstract}
We explore the properties of a recently proposed background independent exact renormalization group approach to gauge theories and gravity. In the process we also develop the machinery needed to study it rigorously. The proposal comes with some advantages. It preserves gauge invariance manifestly, avoids introducing unphysical fields, such as ghosts and Pauli-Villars fields, and does not require gauge-fixing. However, we show that in the simple case of $SU(N)$ Yang-Mills it does not completely regularise the longitudinal part of vertex functions already at one loop, invalidating certain methods for extracting universal components. Moreover we demonstrate a kind of no-go theorem: within the proposed structure, whatever choice is made for covariantisation and cutoff profiles, the two-point vertex flow equation at one loop cannot be both transverse, as required by gauge invariance, and fully regularised. 
\end{abstract}


\end{titlepage}

\tableofcontents

\newpage

\section{Introduction}
\label{sec:Introduction}

Understanding the Wilsonian renormalization group (RG) structure of quantum gravity is surely of importance,  see \eg \cite{Stelle:1976gc,Adler:1982ri,Weinberg:1980,Donoghue:1994dn,Reuter:1996,Percacci:2017fkn,Reuter:2019byg,Bonanno:2020bil,Ambjorn:2012jv,Loll:2019rdj}, and central to this is the r\^ole of diffeomorphism invariance. One approach is to try to generalise the exact RG \cite{Wilson:1973,Wegner:1972ih} to gravity, in such a way that it is manifestly diffeomorphism invariant. On the one hand this would allow computations to be done whilst keeping exact diffeomorphism invariance at every stage, \ie without gauge fixing, and on the other hand, these computations would be background independent, that is performed without first choosing the space-time manifold and background metric.

At the classical level this can be done \cite{Morris:2016nda}. However in order to compute quantum corrections, extra ultraviolet regularisation has to be incorporated into the exact RG so that the integration is properly cut off in some diffeomorphism invariant way at the effective cutoff scale $\Lambda$. For the simpler case of gauge invariance in $SU(N)$ Yang-Mills theory this problem was solved by incorporating gauge invariant PV (Pauli-Villars) fields arising from a spontaneously broken $SU(N|N)$ gauge theory \cite{Morris:1998kz,Morris:1999px,Morris:2000fs,Morris:2000jj,Arnone:2000bv, Arnone:2000qd,Arnone:2001iy,Arnone:2002qi,Arnone:2002fa,Arnone:2002cs,Arnone:2005fb,Arnone:2002fb,Gatti:2002kc,Morris:2005tv,Rosten:2004aw,Rosten:2005qs,Rosten:2005ka,Rosten:2006tk,Rosten:2006qx,Arnone:2006ie,Rosten:2010vm,Rosten:2011ty,Falls:2017nnu}. It is not clear whether one can generalise such a scheme to gravity however, in particular it is not clear what should play the r\^ole of $SU(N|N)$, although a kind of supergravity has been suggested in \cite{Kellett:2020rjw}. 

Recently a different approach to the problem of regularisation has been pursued \cite{Falls:2020tmj}. Explicit PV fields are avoided and instead replaced by functional determinants which, if constructed as squares of simpler determinants, can be shown to work in the framework of standard perturbation theory \cite{Slavnov:1977zf,Bakeyev:1996is,Asorey:1989ha}. Furthermore a geometric approach is followed where the determinants can be regarded as defining a regularised volume element on the orbit space of the gauge theory \cite{Asorey:1989ha}. The flow equation is then formulated in a way that is manifestly invariant under field redefinitions, and applies equally well to both gravity and gauge theory \cite{Falls:2020tmj}.

Although this new proposal has elegant features, it is not immediately evident that the regularisation is successfully implemented in the flow equation once we delve into the details of the relevant Feynman diagrams, first at one loop and then also at higher loops. In this paper we put the proposal to the test by constructing explicit expressions for the relevant vertices  and then carefully analyse their UV (ultraviolet) behaviour, as a function of loop momentum, in the form that they appear in quantum corrections. For this purpose it is sufficient to focus on Yang-Mills since if the regularisation fails for Yang-Mills it most certainly fails for quantum gravity (given the latter's poor UV behaviour). Working with Yang-Mills also means we can take over methods used for these investigations in the earlier successful construction \cite{Morris:1998kz,Morris:1999px,Morris:2000fs,Morris:2000jj,Arnone:2000bv, Arnone:2000qd,Arnone:2001iy,Arnone:2002qi,Arnone:2002fa,Arnone:2002cs,Arnone:2005fb,Arnone:2002fb,Gatti:2002kc,Morris:2005tv,Rosten:2004aw,Rosten:2005qs,Rosten:2005ka,Rosten:2006tk,Rosten:2006qx,Arnone:2006ie,Rosten:2010vm,Rosten:2011ty,Falls:2017nnu}. As we will see, the proposal of ref. \cite{Falls:2020tmj} unfortunately fails to fully regularise already at one loop, but in a rather subtle way, which in particular invalidates powerful techniques previously used to extract universal information \cite{Arnone:2002cs,Arnone:2002yh}.

The paper is organised as follows. In section \ref{sec:Background independent regularisation} we review the proposal, and in the process, set out our notation and conventions. In sec. \ref{sec: regularisation failure sketch} we sketch why the regularisation can fail.
In section \ref{sec:Perturbative expansion} we build the machinery needed to rigorously test the structure of the renormalization scheme at the perturbative level. In sec. \ref{sec:Regularisation of higher point classical vertices} we confirm that the higher point classical vertices incorporate the assumed regularisation. Then in sec. \ref{sec:One loop beta function} we apply the techniques extensively to an analysis of the simplest one-loop correction namely that for the effective action two-point vertex. We show that the regularisation is sufficient for the momentum independent part, giving a vanishing result as it should by gauge invariance, only if the $\Lambda$-derivative of quadratically divergent constant part is discarded. (Recall that $\Lambda$ is the effective cutoff scale.) The part that is second order in momentum ought to give the one-loop beta function, if properly regulated. However we show that the result cannot both be completely regularised and transverse. It can be taken to be transverse only if the $\Lambda$-derivative of a linearly divergent part is discarded. 
Since this holds for all choices of covariantisation and cutoff profiles, it points to some inherent limitations in the structure of the proposed flow equation. In sec. \ref{sec:Summary and Conclusions} we summarise and draw our conclusions.

\section{A proposal for a background independent exact RG}
\label{sec:Background independent regularisation}

\subsection{Notation and basic ingredients for regularisation}
\label{sec:Basic ingredients for regularisation}

The flow equation proposed in ref. \cite{Falls:2020tmj} incorporates a geometric approach to the quantisation of gauge theories \cite{Mottola:1995sj,dewitt2003global} and as such it is useful to use DeWitt notation. On the other hand, as we will see, a detailed understanding of the UV properties can only be reached by working with explicit expressions for the vertices. In the case of Yang-Mills theories, the flow equations then take their simplest form if we regard the gauge fields as valued in the Lie algebra, \ie contracted into the generators \cite{Morris:1998kz,Morris:1999px}. We will therefore work with both notations as appropriate. 

We work in Euclidean signature and, therefore, we will keep all gauge group indices as superscripts and Lorentz indices as subscripts for convenience. For position and momentum space integrals we introduce the shorthand
\beal
\int d^D\! x &= \int_x \,, & \int \frac{d^D \!p}{(2\pi)^D} &= \int_p \,,
\eeal
respectively, where $D$ is the number of dimensions we are working in. We will frequently switch between position and momentum space representations throughout the paper and this will help us keep everything clean and concise. For similar reasons we will adopt the following convention:
\be
    \delta(p) \equiv (2\pi)^D \delta^{(D)}(p) \,,
\ee
where $\delta^{(D)}(p)$ is the standard $D$-dimensional Dirac delta function.
Our convention for Fourier transforms is then
\be
\phi(p) = \int_x\! \phi(x)\, e^{-ipx} \,.
\ee
We will use the DeWitt compact notation and its explicit representation interchangeably, so for example
\be
\phi^a J_a \equiv \sum_a \int_x A_\mu^a(x) J_\mu^a(x)\,.
\ee
Last, but not least, we will adopt the following shorthands for momenta and scalar functions of momenta, respectively: 
\beal
    \frac{p}{\Lambda} &\equiv \Tilde{p} \,, \\
    K\bigg( \frac{p^2}{\Lambda^2} \bigg) &\equiv K_p \,,
\eeal
where $\Lambda$ is the effective cutoff energy scale.

We work with the gauge group SU($N$). We use DeWitt Latin indices from the start of the alphabet to label gauge fields, thus $\phi^a$, but as already mentioned, when we need more explicit expressions it will be convenient to regard the gauge fields $A_\mu(x)$ as contracted into the generators:
\be
\phi^a \equiv A_\mu(x) = A_\mu^a(x) T^a \,.
\ee
With appropriate definitions for the vertex functions in either language, the expressions will of course be equal, and the two representations are thus equivalent.
The generators $T^a$ are taken to be hermitian, in the fundamental representation, and orthonormalized as $tr(T^a T^b) = \frac{1}{2} \delta^{ab}$. A second set of DeWitt Greek indices from the start of the alphabet is used to label gauge parameters; the map from the two languages is thus:
\be
\epsilon^\alpha \equiv \omega(x) = \omega^a(x) T^a \,.  
\ee

If we adopt the geometric approach of \cite{Mottola:1995sj,dewitt2003global} to the quantisation of gauge theories, we regard the fields $\phi^a$ as coordinates on  an infinite dimensional `manifold' $\Phi$, the space of all possible field configurations. This has the structure of a fibre bundle, and the fibres are the gauge orbits $\mathcal{G}$. However, all the physics happens on the quotient space $\Phi/\mathcal{G}$, where each point belongs to a unique equivalence class $\{\phi\}$, which encompasses all the possible field configurations related by gauge transformations. The generators of the gauge transformation are of the following form:
\be
\label{generators}
    K_{\alpha}^{a}[\phi] \equiv D_\mu \delta(x-y) \,,
\ee
where the covariant derivative is given by $D_\mu := \partial_\mu - i A_\mu(x)$, and $A_\mu$ is understood to act by commutation. This means that gauge transformations can be written equivalently as
\beal
    \delta \phi^{a} &= K_{\alpha}^{a}[\phi] \: \epsilon^{\alpha} \,,\\
    \delta A_\mu(x) &= D_\mu \cdot \omega := [D_\mu,\omega] \,. \label{gaugetransf}
\eeal
The field strength is given by $F_{\mu\nu} := i [D_\mu,D_\nu]$. We write the SU($N$) Yang-Mills action in the following way:
\be
\label{YM}
        I[\phi] \equiv \frac{1}{4g^2} \int_x (F^a_{\mu\nu})^2 = \frac{1}{2g^2} \: \text{tr}\int_x F_{\mu\nu}^2 \,.
\ee
In momentum space this becomes:
\besp
        I[\phi] \equiv \frac{1}{2 g^2} \int_p I_{\mu_1\mu_2}(p) \: \text{tr}\Big(A_{\mu_1}(p) A_{\mu_2}(-p)\Big) \: + \\
         + \frac{1}{3 g^2} \int_{p_1 \ldots p_3} I_{\mu_1 \ldots \mu_3}(p_1 \ldots p_3) \: \text{tr}\Big(A_{\mu_1}(p_1) \ldots A_{\mu_3}(p_3)\Big) \: \delta(p_1 + p_2 + p_3) \: + \\
         + \frac{1}{4 g^2} \int_{p_1 \ldots p_4} I_{\mu_1 \dots \mu_4}(p_1 \ldots p_4) \: \text{tr}\Big(A_{\mu_1}(p_1) \ldots A_{\mu_4}(p_4)\Big) \: \delta(p_1 + \cdots + p_4) \,,
\eesp
where the vertex functions $I^{(2)}, I^{(3)}$ and $I^{(4)}$, are given by:
\beal
\label{I2}
    I_{\mu_1\mu_2}(p)  &= 2 \, \square_{\mu_1\mu_2}(p) := 2(p^2 \delta_{\mu_1\mu_2} - p_{\mu_1} p_{\mu_2}) \,, \\
    \label{I3}
    I_{\mu_1 \ldots \mu_3}(p_1 \ldots p_3) &= 2 ({p_3}_{\mu_2} \delta_{\mu_1\mu_3} - {p_3}_{\mu_1} \delta_{\mu_2\mu_3}) + \text{cycles} \,, \\
    \label{I4}
    I_{\mu_1 \dots \mu_4}(p_1 \ldots p_4) &= (\delta_{\mu_1\mu_4} \delta_{\mu_2\mu_3} - \delta_{\mu_1\mu_3} \delta_{\mu_2\mu_4}) + \text{cycles} \,.
\eeal
We note that `cycles' stands for all cyclic permutations of momenta ${p_1}_{\mu_1} \rightarrow {p_2}_{\mu_2} \rightarrow \ldots \rightarrow {p_1}_{\mu_1}$. The exact preservation of gauge invariance at all stages during the flow, together with the choice to rescale the coupling in front of the integral in \eqref{YM}, ensures that $A_\mu$ cannot run: any wavefunction renormalization by $Z \neq 1$, $A_\mu^R = Z^{-1/2} A_\mu$, would break gauge invariance: $\delta A_\mu^R = Z^{-1/2} \partial_\mu \omega - i [A_\mu^R,\omega]$ \cite{Abbott:1981ke,Morris:1998kz,Morris:1999px}. Therefore the coupling $g = g(\Lambda)$ is the only quantity that runs. 

Functional derivatives are written as
\be
\label{functional}
    S,_a = \frac{\delta S}{\delta \phi^a} \equiv  \frac{\delta S}{\delta A_\mu(x)} \,,
\ee
where $\delta/\delta A_\mu(x) := 2 \: T^a \: \delta/\delta A_\mu^a(x)$. The properties of the latter can be understood as follows \cite{Morris:1999px,Morris:2000fs}. For convenience here we temporarily write the gauge fields $A_\mu(x)$ simply as $A$, suppressing spacetime dependence and Lorentz indices. Given a well-behaved gauge invariant function $f(A)$ such that $\delta f(A) = \text{tr}(\delta A \, X)$, for some $X$, we can exploit the $SU(N)$ completeness relation to see that:
\be
\frac{\delta f}{\delta A} = X - \frac{1}{N} \text{tr} X \,,
\ee
effectively isolating $X$. This property will be crucial in our endeavour later on. Following the same reasoning, we have two more useful relations:
\be
\label{sowing}
\text{tr} \, Y \frac{\delta f}{\delta A} = \text{tr} YX - \frac{1}{N} \text{tr} Y \text{tr} X \,,
\ee
and, 
\be
\label{splitting}
\text{tr} \frac{\delta}{\delta A} W = \text{tr} Y \text{tr} Z - \frac{1}{N} \text{tr} Y Z \,,
\ee
for some $W = T^a W^a$ such that $\delta W = Y \, \delta A \, Z$.
The functional derivatives of $I[\phi]$ are (in position space):
\beal
    \mathcal{E}_a = I,_a &\equiv \frac{1}{g^2} D_\alpha F_{\mu\alpha} \,,\label{E} \\
    I,_{ab} &\equiv \frac{1}{g^2} \Big( i F_{\mu\nu} + D_\nu D_\mu - \delta_{\mu\nu} D^2 \Big) \delta(x-y) \,.
\eeal

Ultra-local metrics are introduced on  $\Phi$ and $\mathcal{G}$, respectively, such that their corresponding line element (\eg $\delta \phi^a \gamma_{ab} \delta \phi^b$) is made dimensionless by using the appropriate power of the effective scale $\Lambda$ \cite{Falls:2020tmj}. They allow us  to raise or lower DeWitt indices, \eg $\phi^a = \gamma^{ab} \phi_b$. At this abstract level the formulation can be developed in a way that applies equally well to gravity, and indeed ref. \cite{Falls:2020tmj} treats gravity as a special case. For 
gravity however, $\gamma_{ab}$ and $\eta^{\alpha\beta}$ necessarily depend on $\phi$ (which in this case is the spacetime metric). Then to keep equations covariant on $\Phi$ requires the use of a connection $\Gamma^c_{ab}$ in function space and covariant derivatives, \eg replacing $I_{,ab}$ above with $I_{;ab} = I_{,ab}-\Gamma^c_{ab}I_{,c}$. Since we will treat only Yang-Mills theory in this paper, where one can take the metrics to be $\phi$ independent:
\beal
    \gamma_{ab} &= \frac{\Lambda^2}{g^2} \delta_{\mu \nu} \delta(x - y) \,, & \gamma^{ab} &= \frac{g^2}{\Lambda^2} \delta_{\mu \nu} \delta(x - y) \,, \\
    \eta_{\alpha \beta} &= \frac{\Lambda^4}{g^2} \delta(x - y) \,, & \eta^{\alpha \beta} &= \frac{g^2}{\Lambda^4} \delta(x - y) \,,
\eeal
we do not need this extra complication and thus we work only in the flat function space limit.

Again
following \cite{Falls:2020tmj}, we will introduce higher covariant derivatives into the effective action $S$ via functions of Laplace-like differential operators, which we list below:
\beal
\label{Delta}
    {\Delta^a}_b &= \gamma^{ac} I,_{cb} + K^a_\alpha \eta^{\alpha \beta} K^c_\beta \gamma_{cb} & &\equiv & \Delta_{\mu\nu} &= \Lambda^{-2}\left( 2 i F_{\mu\nu} - \delta_{\mu\nu} D^2 \right) \delta(x-y) \,, \\
\label{Deltapar}
    {\big(\Delta_\parallel \big)^\alpha}_\beta &= \eta^{\alpha\delta} K^a_\delta \gamma_{ab} K^b_\beta & &\equiv & \Delta_\parallel &= -\Lambda^{-2} D^2 \delta(x-y) \,.
\eeal
For two (matrix valued) functions $J_\mu(x)$ and $J_\nu(x)$ and a momentum kernel $K_{\mu\nu}(p^2/\Lambda^2)$ we introduce the shorthand:
\beal
\label{K}
J_{\mu} \cdot K_{\mu\nu} \cdot J_{\nu} &:= 2 \, \text{tr} \int_x J_\mu(x) K_{\mu\nu}\left( -\partial^2/\Lambda^2 \right) J_\nu(x) \,, \\
K_{\mu\nu} \left( -\partial^2/\Lambda^2 \right) \delta(x-y) &= \int_p K_{\mu\nu}(p^2/\Lambda^2) \, e^{ip(x-y)} \,.
\eeal

Gauge invariant quantities have all DeWitt indices contracted (\ie behaves as a scalar on $\Phi$ and $\mathcal{G}$). One gauge invariant object that plays a fundamental role in the regularisation is the action $A$ :
\be
    A = I + \frac{1}{2} I,_a \big[ X(\Delta) \big]^{ab} I,_b \label{A} \,.
\ee
It is \emph{quasi-local}, \ie has an all orders expansion in momenta \cite{Morris:1999px,Morris:2000fs}.
The kernel $X(z)$ is given by:
\be
\label{X}
    X(z) = \frac{1 - c(z)}{z  c(z)} \,,
\ee
where $c(z)$ is a smooth UV cutoff profile such that $c(0) = 1$, and $c(p^2/\Lambda^2) \rightarrow 0$ as $p^2/\Lambda^2 \rightarrow \infty$.

The anomalous dimension is $\eta := - g^2 \Lambda \partial_\Lambda (1/g^2) = \frac{2}{g}\Lambda\partial_\Lambda g$. Given that $\eta \propto \hbar$ (\ie actually $\hbar g^2$), we can write down a loop expansion for it:
\be
    \eta = \eta_1 g^2 + \eta_2 g^4 + \cdots \,,
\ee
and thus for $\beta := \Lambda \partial_\Lambda g$ as well ($\eta_i=2\beta_i$):
\be
    \beta = \beta_1 g^3 + \beta_2 g^5 + \cdots \,.
\ee

It will prove useful to analyse expressions using the following projection operators:
\beal
    {\Pi_T}^a{}_b &:= \delta^a{}_b - {\Pi_L}^a{}_b &&\equiv &\Pi^T_{\mu\nu} &:= \delta_{\mu\nu} - \Pi^L_{\mu\nu} \,, \\
    {\Pi_L}^a{}_b &:= K^a_\alpha \left( \Delta_\parallel^{-1} \right)^{\alpha\beta} K^c_{\beta} \gamma_{bc} &&\equiv &\Pi^L_{\mu\nu} &:= D_{\mu} \Big( D^2 \Big)^{-1} D_{\nu} \,,
\eeal
although they are not directly involved in the regularisation scheme because of their non-local nature.

\subsection{The main idea}
\label{sec:The main idea}

Before delving into details, it is instructive to outline the main steps undertaken in ref.\cite{Falls:2020tmj} to construct a flow equation for the effective action $S$. Essentially this is a three step procedure, where the first two steps represent a geometric reformulation of Slavnov's higher derivative regularisation scheme \cite{Bakeyev:1996is,Slavnov:1977zf}, while in the last step the freedom one has to design an exact RG \cite{WR,Morris:1998kz,Latorre:2000qc,Latorre:2000jp} is exploited to avoid fixing the gauge and to manifestly preserve gauge invariance at all stages during the flow. 

First of all we need to incorporate higher covariant derivatives, $S_{UV}$, into the effective action $S$, to improve the UV behaviour by modifying the propagators:\footnote{As it stands, this equation is almost devoid of meaning since $\mathcal{S}$ could modify or even cancel terms in $S_{UV}$. At the classical level, we give it meaning  by insisting that $\mathcal{S}^0$ contains only three-point and higher vertices:  $\mathcal{S}^{0(n<3)}=0$. , which, in turn, fixes the on-shell two-point classical effective action. This choice will allow us to compute classical vertices in terms of quantities already known \cite{Arnone:2002cs}.}
\be
\label{calSdef}
    S = I + S_{UV} + \mathcal{S}\,, 
\ee
where $\mathcal{S}$ stands for all the other interactions in the effective action that are generated by the flow.
However higher covariant derivatives do not regularise all one-loop divergences, and therefore further regularisation is required \cite{Bakeyev:1996is,Slavnov:1977zf}.

The second step addresses this issue by introducing PV regularisation, which however is introduced directly through functional determinants in such a way that these act as \emph{quasi-local} metrics for $\Phi$, and $\mathcal{G}$; they are called $G_{ab}[\phi]$ and $H_{\alpha\beta}[\phi]$ respectively. Now, the partition function $\mathcal{Z}$ takes in general the following form:
\be
    \mathcal{Z} = \int_{\Phi / \mathcal{G}} d\{ \phi \} \: M[\{ \phi \}] \: e^{-S[\{ \phi \}]} \,,
\ee
where $M[\{\phi\}]$ is the volume element on the quotient space $\Phi/\mathcal{G}$. Factoring the integration measure using the functionals mentioned above,   
\be
    d\phi \sqrt{det \: G[\phi]} = d\{ \phi \} M[\{ \phi \}] \: d\xi \sqrt{det \: H[\phi]} \,,
\ee
we thus rewrite the partition function (Tr stands for a functional trace of the kernels):
\be
\label{partition}
    \mathcal{Z} \propto \int_{\Phi} d\phi \: \frac{\sqrt{det \: G[\phi]}}{\sqrt{det \: H[\phi]}} \: e^{-S[\phi]} = \int_{\Phi} d\phi \: e^{- S[\phi] + \frac{1}{2} \text{Tr} \ln G_{ab}[\phi] - \frac{1}{2} \text{Tr} \ln H_{\alpha\beta}[\phi]} 
\ee
(up to some actually-infinite constant that we are free to discard without altering the physics).

The idea then is that $G$ is chosen to cancel one-loop UV divergences arising from $S$. In the infrared (IR), $G$ and $H$ are both order $\Lambda^2$, corresponding to PV masses of order the cutoff. $G$ thus has a longitudinal part which will contribute divergences, but $H$ can be used to cancel them. Finally by choosing $G$ and $H$ each such that they factor into other operators that are less divergent, we can avoid so-called `overlapping' divergences, ones that correspond at one loop to having external PV legs \cite{Faddeev:1980be,Bakeyev:1996is,Asorey:1995tq}. Thus at this stage, PV regularisation will require the following: 
\begin{itemize}
    \item [(PV1)]At high momentum the Hessian for fluctuations matches that of $G$ in the transverse space:
    \be
        {\Pi_T}^c{}_a \, G_{cd} \, {\Pi_T}^d{}_b \thicksim S_{,ab} \,.
    \ee
    
    \item [(PV2)] Similarly $H$ matches the longitudinal part of $G$:
    \be
        K^a_\alpha G_{ab} K^b_\beta \thicksim H_{\alpha\beta} \,.
    \ee
    
    \item [(PV3)]$G_{ab}[\phi]$ and $H_{\alpha\beta}[\phi]$ must be chosen so that they factor into operators that are less divergent.
\end{itemize}

Slavnov's original scheme does not avoid fixing the gauge \cite{Bakeyev:1996is,Slavnov:1977zf}. In ref.\cite{Falls:2020tmj} gauge fixing is completely avoided by exploiting the freedom \cite{WR,Morris:1998kz,Morris:1999px,Latorre:2000qc,Latorre:2000jp,Arnone:2002yh,Arnone:2003pa} one has to design an exact RG \cite{Wilson:1973,Wegner:1972ih} flow for the effective action $S$. This leads to the final flow equation:
\be
\label{flowequation}
    \Lambda D_\Lambda S = \frac{1}{2} \text{Tr} \big[ G^{-1} \Lambda D_\Lambda G - H^{-1} \Lambda D_\Lambda H \big] \,,
\ee
where $\Lambda D_\Lambda := \Lambda \partial_\Lambda + \mathcal{L}_\Psi$ is the total RG derivative, $\mathcal{L}_\Psi$ being a Lie derivative on $\Phi$ associated to the change of variables $\phi^a\mapsto\phi^a-\delta\Lambda\Psi^a$.  The total RG derivative acts in the following way:
\beal
    \Lambda D_\Lambda S &= \Lambda \partial_\Lambda S + \Psi^a S,_a \,, \label{StotalRG}\\
    \Lambda D_\Lambda H_{\alpha\beta} &= \Lambda \partial_\Lambda H_{\alpha\beta} + \Psi^a H_{\alpha\beta,a} \,, \label{HtotalRG}\\
    \Lambda D_\Lambda G_{ab} &= \Lambda \partial_\Lambda G_{ab} + \Psi^c G_{ab,c} + {\Psi^c}_{,a} G_{bc} + {\Psi^c}_{,b} G_{ac} \label{GtotalRG}\:.
\eeal

The strategy is that, altogether, the higher covariant derivatives and PV field determinants should provide a regularised `kinetic' $I+S_{UV} \in S$ part, while a careful choice of $\Psi^a$ is meant to ensure that the exact RG then generates only an $\mathcal{S} \in S$ part that is free of divergences.

\subsection{Regularisation structure}
\label{sec:Regularisation structure}

The method proposed in ref.\cite{Falls:2020tmj} to construct the regularisation scheme, \ie PV operators and higher covariant derivatives, proceeds as follows. First of all, we need to construct $G_{ab}[\phi]$ and $H_{\alpha\beta}[\phi]$ by taking into account the requirements outlined in section \ref{sec:The main idea}. To satisfy (PV3) they are set to
\beal
    G_{ab} &= {\big( C^{-1} \big)^c}_a \: \gamma_{cd} \: {\big( C^{-1} \big)^d}_b \,, \label{Gdef}\\ H_{\alpha\beta} &= {B^\gamma}_\alpha \: b_{\gamma\delta} \: {B^\delta}_\beta \,, \label{Hdef}
\eeal
where we require $B^\alpha{}_\beta$, $b^\alpha{}_\beta$, and $\big( C^{-1} \big)^a{}_b$ to be constructed such that $G_{ab} \rightarrow \gamma_{ab}$ and $H_{\alpha\beta} \rightarrow \eta_{\alpha\beta}$ as $\Lambda \rightarrow \infty$. This latter requirement ensures that in the continuum limit the \emph{quasi-local} and \emph{ultra-local} metrics coincide, and hence encode the same physics \cite{Rosten:2011ty}. Furthermore, the condition (PV1) implies that $C^2 S^{(2)} \thicksim 1$, which essentially means that all divergences coming from $C^{-1}$ vertices should be regulated by contributions in the effective action involving the effective propagator. The remaining condition (PV2) together with \eqref{Gdef}, and \eqref{Hdef} relates the PV operators to one another in the high momentum limit:
\beal
    \big( C^{-1} \big)^a{}_b K^b_\alpha &\thicksim K^a_\beta B^\beta{}_\alpha \,, \label{PV2a}\\
    b_{\alpha\beta} &\thicksim \big( \Delta_\parallel \big)_{\alpha\beta} = K^a_\alpha \gamma_{ab} K^b_\beta \label{PV2b} \,.
\eeal

The second part of the regularisation scheme amounts to introducing higher covariant derivatives into the effective action $S$. Given that $G$ is bi-linear in $C^{-1}$ we will keep the same structure for $S_{UV}$ \cite{Bakeyev:1996is},
\be
\label{SUV}
    S_{UV} = \frac{1}{2} A_{,a} \gamma^{ab} A_{,b} \,,
\ee
and thus write the effective action $S$ as 
\be
\label{S}
    S = I + \mathcal{S} + \frac{1}{2} A_{,a} \gamma^{ab} A_{,b} \,.
\ee
Here it is instructive to pause and make some comments on the choice of using $A$ and on the strategy for analysis of the regularisation and flow equation. If we functionally differentiate \eqref{A} twice we get:
\be
\label{2ptA}
    A_{,ab} = \Big( \Delta \cdot c^{-1}\big( \Delta \big) \Big)_{ad} {\Pi_T}^d{}_b + O\big( \mathcal{E} \big) \,.
\ee
Following ref. \cite{Falls:2020tmj} it is useful to write equations such as this where the equality is given up to terms that vanish on the equations of motion $\mathcal{E}$, \cf \eqref{E}.
Since $\mathcal{E}$ vanishes if we set $A_\mu=0$, we can read off from the above that the first term  provides the effective inverse propagator that would arise from using $A$ as the action. We see that it is transverse and that it would lead to an effective propagator $\thicksim c_q/q^2$ which is regularised by a cutoff function in the usual way. The first term above is also solely responsible for the one-loop contribution that $A$ would provide (through the functional determinant of $ A_{,ab}$) when the result is evaluated on shell.  From \eqref{2ptA} we can also infer the large momentum behaviour of the vertices $A^{(n)}$. Clearly $A^{(2)} \thicksim q^2 c_q^{-1}$. In fact, for any integer $n$, $A^{(n)} \thicksim c_q^{-1}$ up to some multiplicative power of momentum which can be ignored in comparison if the cutoff is chosen to be strong enough, for example $c(x) = e^{-x}$ or $c(x) = (1+x)^{-m}$ for some large $m$.  To write this we will use the following shorthand $A^{(n)} \approx c_q^{-1}$. These properties will be more precisely defined and confirmed in sec. \ref{sec:Large momentum behaviour}. Until then in the spirit of ref. \cite{Falls:2020tmj}, we count only powers of cutoffs to analyse the behaviour of different expressions in the large momentum limit. For example, looking at \eqref{S} we see that
\be
    S_{,ab} \thicksim A^{(2)} A^{(2)} \approx c_q^{-2} \,,
\ee
which means that the (PV1) condition can now be recast as:
\be
\label{PV1}
    C^{-1} \approx c_q^{-1} \,. 
\ee

To properly implement regularisation, (PV1-3) ought to be fulfilled or, equivalently, \eqref{PV2a}, \eqref{PV2b} and \eqref{PV1}. A particular choice which satisfies this is the following :
\beal
    {\big( C^{-1} \big)^a}_b &= {\delta^a}_b + \gamma^{ac} A,_{cb} + K^a_\alpha Y^{\alpha\beta} K^c_\beta \gamma_{cb} \,, \label{C^-1} \\
    {B^\alpha}_\beta &= {\delta^\alpha}_\beta + Y^{\alpha\gamma} \big( \Delta_\parallel \big)_{\gamma\beta} \,, \label{B} \\
    {b^\alpha}_\beta &= {t^\alpha}_\beta + {\big( \Delta_\parallel \big)^\alpha}_\beta \label{b} \,,
\eeal
where $Y = Y \big( \Delta_\parallel \big)$ is a quasi-local function such that $Y(z) \rightarrow \infty$ as $z \rightarrow \infty$ at the same rate as $1/c(z)$ and $Y(0)$ is finite, while $t = t \big( \Delta_\parallel \big)$ is a quasi-local function such that $t(z) \rightarrow 0$ as $z \rightarrow \infty$ and $t(0) = 1$. 

\subsection{Flow equation}
\label{sec:Flow equation}


For the flow equation to have fixed points it has to be non-linear which implies that the blocking functional $\Psi[\phi]$ must itself depend on $S$. Generalising \cite{Morris:1998kz,Morris:1999px,Morris:2000fs,Morris:2000jj,Arnone:2000bv, Arnone:2000qd,Arnone:2001iy,Arnone:2002yh,Arnone:2002qi,Arnone:2002fa,Arnone:2003pa,Arnone:2002cs,Arnone:2005fb,Arnone:2002fb,Gatti:2002kc,Morris:2005tv,Rosten:2004aw,Rosten:2005qs,Rosten:2005ka,Rosten:2005ep,Rosten:2006tk,Rosten:2006qx,Arnone:2006ie,Rosten:2010vm,Rosten:2011ty,Falls:2017nnu}, which are themselves generalisations of the Polchinski equation \cite{Polchinski:1983gv}, the blocking functional is written in the following way:
\be
\label{blockingfunctionaldef}
    \Psi^a = - \frac{1}{2} \mathcal{K}^{ab} \Sigma_{,b} + \psi^a \,,
\ee
where 
\be
\label{Sigma}
\Sigma := S - \hat{S} = \mathcal{S} - \frac{1}{2} I_{,a} X^{ab} I_{,b}\,,
\ee
and the ``seed'' action $\hat{S}$ is given by:
\be
\label{seed action}
\hat{S} := A + S_{UV} = A + \frac{1}{2} A_{,a} \gamma^{ab} A_{,b} \,.
\ee
We will call the quasi-local functionals $\mathcal{K}^{ab}$  and $\psi^a$ the \textit{exact RG kernels}. We will try to fine-tune them to remove the unwanted divergences still present in the flow equation. In addition to this, it will prove useful to fix the classical two-point function \cite{Morris:1999px,Morris:2000fs,Arnone:2002cs}. As such, we require 
\be
\label{S0}
    S_0 = I + \frac{1}{2} A_{,a} \gamma^{ab} A_{,b} + \mathcal{S}_0 \,,
\ee
where $\mathcal{S}_0 = O \big( \mathcal{E}^3 \big)$, to be a solution of the classical flow equation 
\be
\label{classflow}
\Lambda D_\Lambda S_0 = 0\,.
\ee 
This fixes the classical two-point function to be
\be
    S_{0,ab} = I_{,ab} + A_{,ma} \gamma^{mn} A_{,nb} + O \big( \mathcal{E} \big) \,.
\ee

The only thing left now is to fix the exact RG kernels. The main constraint comes from the left-hand side of \eqref{flowequation}, \ie the classical flow equation. One needs to carefully choose the kernels such that the effective interactions generated during the flow (\ie $\mathcal{S}$) do not destroy the regularisation. In other words we need to ensure that $\mathcal{S}$ diverges more weakly than $S_{UV} \approx c_q^{-2}$. Since we are unable to write a closed form expression for $\mathcal{S}$ (we can only determine a finite number of higher order contributions to it by repeated iterations), following ref. \cite{Falls:2020tmj} we suppose that its maximum divergence rate is given by: 
\be
\label{assumption1}
    \mathcal{S}^{(n)} \thicksim A^{(n)} \approx c_q^{-1} \,,
\ee
or, equivalently, using the definition \eqref{Sigma} of $\Sigma$, that
\be
\label{assumption2}
    \Sigma^{(n)} \approx c_q^{-1} \,.
\ee
The strategy now is to  fine-tune the exact RG kernels such that they cancel all the terms which diverge faster than $c_q^{-1}$ in the flow equation of $\mathcal{S}^{(n)}$, and hence prove \textit{a posteriori} that the assumption \eqref{assumption1} is consistent. If we unwrap the left-hand side of \eqref{flowequation} we get:
\besp
\label{explicitfloweq}
    \Lambda D_\Lambda S = \dot{A}_{,a} \gamma^{ab} A_{,b} + \frac{1}{2} A_{,a} \dot{\gamma}^{ab} A_{,b} + A_{,a} \big( C^{-1} \big)^a{}_b \psi^b - \frac{1}{2} A_{,c} \big( C^{-1} \big)^c{}_a \mathcal{K}^{ab} \Sigma_{,b} \\
    + \dot{A} + \dot{\Sigma} - \frac{1}{2} \Sigma_{,a} \mathcal{K}^{ab} \Sigma_{,b} + \Sigma_{,a} \psi^a \,,
\eesp
where `` $\dot{}$ " denotes the RG time derivative $\Lambda \partial_\Lambda$. To arrive at this expression we have used the identity:
\be
\label{Shatid}
    \hat{S}_{,b} = A_{,a} \big( C^{-1} \big)^a{}_b \,,
\ee
which follows from \eqref{C^-1} and the gauge invariance of $A$, $K^a_\alpha A_{,a} = 0$. The first two terms of \eqref{explicitfloweq} are bi-linear in $A_{,a}$ and hence diverge as $\approx c_q^{-2}$. To eliminate them we use the third term, setting
\be
\label{psikernel}
    \psi^a = - C^a{}_b \Big( \gamma^{bc} \dot{A}_{,c} + \frac{1}{2} \dot{\gamma}^{bc} A_{,c} \Big) \,.
\ee
Now looking at the remaining terms, we see that the potentially dangerous ones are the fourth term which, ignoring $\mathcal{K}^{ab}$ for the moment, diverges as $\approx c_q^{-3}$, and similarly the next to last term, which diverges as $\approx c_q^{-2}$. 
Cancelling the $C^{-1}$ factor in the former by writing
\be
\label{Kkernel}
    \mathcal{K}^{ab} = C^a{}_c [\kappa(\Delta)]^{cb} \,,
\ee
the kernel $\kappa^{ab}$ is then fixed uniquely by substituting $S_0$ from \eqref{S0}, into \eqref{explicitfloweq} and requiring cancellation, \ie requiring the classical flow \eqref{classflow} at $O\big( \mathcal{E}^2 \big)$. The result is:
\be
\label{kappakernel}
    \kappa^a{}_b = \bigg( 2 \frac{ c(\Delta) - 2\Delta c^{\prime}(\Delta)}{c^{2}(\Delta)+\Delta} \bigg)^{a}{}_b 
\ee
(see sec. IIIB and app. A of ref. \cite{Falls:2020tmj}).
This means that $\kappa^{ab} \approx c_q$ and hence ref. \cite{Falls:2020tmj} concluded that all the potentially offending terms in \eqref{explicitfloweq} are now brought under control. This part of the flow equation (the classical part)  now simplifies to:
\be
\label{flowequation2}
    \Lambda D_\Lambda S =  - \frac{1}{2} \big( \Sigma_{,a} + \hat{S}_{,a} \big) C^a{}_c \kappa^{cb} \Sigma_{,b}  + \dot{A} + \dot{\Sigma} + \Sigma_{,a} \psi^a \,,
\ee
and indeed a direct substitution of the above estimates suggests that now no terms 
diverge faster than $c_q^{-1}$. Moreover, although more involved, one can similarly check  that the exact RG kernels \eqref{psikernel}, \eqref{Kkernel} and \eqref{kappakernel} ensure that the RHS of \eqref{flowequation} is also apparently properly regularised \cite{Falls:2020tmj}.

\section{Regularisation failure: a sketch}
\label{sec: regularisation failure sketch}

The problem with the analysis reviewed in the previous section is more easily grasped if we convert the analysis into the language of Feynman graphs. 
The reason this is helpful, is because we can then derive closed formulae for all the component vertices and propagator-like terms of these graphs, and then rigorously characterise their asymptotic behaviour in the limit that certain momenta diverge. In particular in quantum corrections, UV divergences arise from the limit in which loop momenta $q\to\infty$ while external momenta are held fixed. The components in the flow equation \eqref{flowequation2} do not necessarily behave in this limit with the assumed powers of $c_q$ because their UV behaviour depends on the details of how $q$ is routed through the vertices.

By focusing in sec. \ref{sec:One loop beta function} on the part of the one-loop action that should provide the beta function, we show that regularisation fails here in a somewhat subtle way. There are parts that are UV divergent, but can be set to zero once treated as $\Lambda$ independent. However, gauge transformations still map the one-loop beta function contribution into those improperly regularised parts. In fact we show that the contribution cannot then both be gauge invariant and properly regularised. Actually this latter conclusion holds for all choices of cutoff functions and some generalisations of the construction. Although the flow equation is constructed to be covariant by using functions of the differential operators $\Delta$ and $\Delta_\parallel$, other choices of covariantisation are possible that still reduce to the same functions when the gauge field $A_\mu$ is set to zero (for example by allowing the functions to depend separately on $F_{\mu\nu}$). Our derivation is independent of these details.

In fact the first hint of regularisation failure seems to come from how the regularisation is constructed in the first place. Ref.\cite{Falls:2020tmj} argues that in the high momentum limit the leading divergent part of the propagator is its transverse part. The argument goes as follows. Starting from the gauge invariance of $S$ (under a gauge transformation $\phi^a \rightarrow \phi^a + \delta\phi^a$),
\be
    K^a_\alpha S_{,a} = 0 \,,
\ee
and differentiating it once we obtain:
\be
\label{DeWitt Ward identity}
   K^a_\alpha S_{,ab} = - K^a_{\alpha,b} S_{,a} \,.
\ee
Regarding this expression as a differential operator one can think of the large momentum as going from $\alpha$ to $b$. Then this equation tells us that the behaviour at large momentum is given by $K^a_{\alpha,b}$, and since $K^a_\alpha$ is a first order differential operator the divergence is only with one power of momentum. However as an expression about vertices more generally we see that we cannot exclude the possibility that the loop momentum threads instead through $S_{,a}$ whose high momentum behaviour can behave as badly as does $S_{,ab}$.

\section{Perturbative expansion}
\label{sec:Perturbative expansion}

In summary, the robustness of the regularisation scheme outlined in sec. \ref{sec:Background independent regularisation} needs to be thoroughly tested by analysing carefully the high momentum behaviour of $n$-point vertices at each order in the loop expansion. In what follows we will lay out the machinery needed to do this properly, before returning to these issues in sec. \ref{sec:One loop beta function}.

\subsection{Loop expansion}
\label{sec:Loop expansion}

The effective action $S$, and in fact any gauge invariant action appearing in this analysis, has a weak coupling expansion (\ie loop expansion) in $g^2$ (actually in $\hbar g^2$):
\be
\label{S loop expansion}
    S = \frac{1}{g^2} S_0 + S_1 + g^2 S_2 + \cdots \,.
\ee
This means that purely classical actions such as $S_0$, but also \eg $A$, $S_{UV}$ and $\hat{S}$, carry only a single $1/g^2$ prefactor. 

The form of the ultralocal metrics $\gamma_{ab}$ and $\eta_{\alpha\beta}$ ensure that the factors of $g$ embedded in the kernels combine with actions to preserve this property. Therefore we simplify matters from now on by ignoring the factors of $g$ entirely and trusting that they can be put back in the form \eqref{S loop expansion} above. (Actually, one can also simplify matters by recognising that the powers of $\Lambda$ appearing through these metrics are as required to make the equations dimensionally correct, and thus essentially ignore these metrics.) Note that in the ensuing we will also place the loop index as a superscript when convenient (so write $S^0\equiv S_0$ \etc).

Similarly to \eqref{S loop expansion} we can write down a loop expansion for the blocking functional $\Psi^a$:
\beal
    \Psi^a &= \Psi_0^a + g^2 \Psi_1^a + \cdots \\
           &= \Big[ -\frac{1}{2} \mathcal{K}^{ab} {\Sigma^0}_{,b} + \psi_0^a \Big] + g^2 \Big[-\frac{1}{2} \mathcal{K}^{ab} {\Sigma^1}_{,b} + \psi^a_1 \Big] + \cdots \label{unrefinedexpansion} \,.
\eeal
This can be further refined. If we explicitly compute the RG time derivatives in \eqref{psikernel} and use the commutativity of RG time and functional derivatives, we can recast $\psi^a$ in the following way:
\be
    \psi^a = C^{ab} \theta_{,b} \,,
\ee
where we have introduced the gauge invariant action $\theta$, given by
\be
\label{theta}
    \theta := \frac{\eta}{2} A + I + I_{,m} [W(\Delta)]^{mn} I_{,n} \,,
\ee
with $W(z) := \frac{3}{2} X(z) + z X'(z)$, which thus has a loop expansion:
\beal
    &\theta^0 = I + I_{,m} W^{mn} I_{,n} \,, \label{theta0} \\
    &\theta^n = \frac{\eta_n}{2} A \,.\label{thetan}
\eeal
This enables us to rewrite \eqref{unrefinedexpansion} as
\be
\label{psi loop expansion}
    \Psi^a = \Big[ -\frac{1}{2} \mathcal{K}^{ab} {\Sigma^0}_{,b} + C^{ab} {\theta^0}_{,b} \Big] + g^2 \Big[-\frac{1}{2} \mathcal{K}^{ab} {\Sigma^1}_{,b} + C^{ab} {\theta^1}_{,b} \Big] + \cdots \,.
\ee

If we substitute \eqref{S loop expansion} and \eqref{psi loop expansion} into the flow equation \eqref{flowequation}, we will obtain its loopwise expansion. Thus we find that the flow of the effective action at the classical level is given by:
\be
\label{0loop}
    \Lambda \partial_\Lambda S^0 = \frac{1}{2} {S^0}_{,a} \mathcal{K}^{ab} {\Sigma^0}_{,b} - {S^0}_{,a} C^{ab} {\theta^0}_{,b} \,,
\ee
at the one-loop level by:
\beal
\label{1loopA}
     \Lambda \partial_\Lambda S^1 &= \eta_1 S^0 + \frac{1}{2} {S^1}_{,a} \mathcal{K}^{ab} {\Sigma^0}_{,b} + \frac{1}{2} {S^0}_{,a} \mathcal{K}^{ab} {\Sigma^1}_{,b} - {S^1}_{,a} C^{ab} {\theta^0}_{,b} - {S^0}_{,a} C^{ab} {\theta^1}_{,b} \nonumber\\ 
     &+ \text{Tr} \Bigg[ C^a{}_c \Big[ \Lambda \partial_\Lambda ( C^{-1} )^c{}_b \Big] + C^a{}_c ( C^{-1} )^c{}_{b,d} \: \Psi_0^d + \frac{\delta \Psi_0^a}{\delta \phi^b} + \delta^a{}_b - (B^{-1})^\alpha{}_\gamma \big[ \Lambda \partial_\Lambda B^\gamma{}_\beta \big] \nonumber\\ 
     &&\hidewidth- (B^{-1})^\alpha{}_\gamma B^\gamma{}_{\beta,a} \Psi_0^a - \frac{1}{2} (b^{-1})^\alpha{}_\gamma \big[ \Lambda \partial_\Lambda b^\gamma{}_\beta \big] -  \frac{1}{2} (b^{-1})^\alpha{}_{\gamma} b^\gamma{}_{\beta,a} \Psi_0^a - 2 \: \delta^\alpha{}_\beta \Bigg] \,, \nonumber\\
     &
\eeal
at the two-loop level by:
\beal
\label{2loop}
     \Lambda \partial_\Lambda S^2 &= \eta_2 S^0 + \frac{1}{2} {S^2}_{,a} \mathcal{K}^{ab} {\Sigma^0}_{,b} + \frac{1}{2} {S^1}_{,a} \mathcal{K}^{ab} {\Sigma^1}_{,b} + \frac{1}{2} {S^0}_{,a} \mathcal{K}^{ab} {\Sigma^2}_{,b} - {S^2}_{,a} C^{ab} {\theta^0}_{,b}  \nonumber\\
     &- {S^1}_{,a} C^{ab} {\theta^1}_{,b} - {S^0}_{,a} C^{ab} {\theta^2}_{,b} + \text{Tr} \Bigg[ C^a{}_c \: ( C^{-1} )^c{}_{b,d} \Psi_1^d + \frac{\delta \Psi_1^a}{\delta \phi^b} - \frac{\eta_1}{2} \delta^a{}_b + \frac{\eta_1}{2} \delta^\alpha{}_\beta \nonumber\\
     &&\hidewidth - (B^{-1})^\alpha{}_\gamma B^\gamma{}_{\beta,a} \Psi_1^a - \frac{1}{2} (b^{-1})^\alpha{}_{\gamma} b^\gamma{}_{\beta,a} \Psi_1^a  \Bigg] \,, \nonumber \\
     &
\eeal
\etc It will prove useful in the subsequent analysis to rewrite \eqref{1loopA} in a different way:
\beal
\label{1loopB}
     \Lambda \partial_\Lambda S^1 &= \eta_1 S^0 + \frac{1}{2} {S^1}_{,a} \mathcal{K}^{ab} {\Sigma^0}_{,b} + \frac{1}{2} {S^0}_{,a} \mathcal{K}^{ab} {\Sigma^1}_{,b} - {S^1}_{,a} C^{ab} {\theta^0}_{,b} - {S^0}_{,a} C^{ab} {\theta^1}_{,b} \nonumber\\ 
     &+ \text{Tr} \Bigg[ \Lambda \partial_\Lambda \big[ \ln \big( C^{-1} \big) \big]^a{}_b + \Psi_0^d \frac{\delta}{\delta \phi^d} \big[ \ln \big( C^{-1} \big) \big]^a{}_b + \frac{\delta \Psi_0^a}{\delta \phi^b} - \Lambda \partial_\Lambda \big[ \ln  B^\alpha{}_\beta \big] \nonumber\\ 
     &&\hidewidth - \Psi_0^a \frac{\delta}{\delta \psi^a} \big[ \ln  B^\alpha{}_\beta \big] - \frac{1}{2} \Lambda \partial_\Lambda \big[ \ln  b^\alpha{}_\beta \big] -  \frac{1}{2} \Psi_0^a \frac{\delta}{\delta \psi^a} \big[ \ln  b^\alpha{}_\beta \big] \Bigg] \,. \nonumber\\
     &
\eeal
Here note that we have also discarded the two vacuum contributions $\delta^a{}_b$ and $\delta^\alpha{}_\beta$.

\subsection{Vertex expansion}
\label{sec:Vertex expansion}

In the above equations we have actions ($S^i,\Sigma^i,\theta^i$) and kernels (\eg $\mathcal{K}^{ab},C^{ab},B^\gamma_\beta$ \etc). To analyse their behaviour precisely we break them down into vertices. The above equations then tell us how these vertices fit together to determine effective action vertices at each loop order. In this subsection we sketch the form of these vertices and their properties \cite{Morris:1999px,Morris:2000fs}.

\subsubsection{Action vertex properties}
\label{sec:Action vertices}

Any gauge invariant action has an expansion in traces  and products of traces \cite{Morris:1999px,Morris:2000fs}, which, in position space, takes the following form:
\begin{flalign}
\label{S trace expansion}
    S =& \sum_{n=2}^\infty \frac{1}{n} \int_{x_1 \ldots x_n} S_{\mu_1 \ldots \mu_n}(x_1 \ldots x_n) \text{tr} \Big( A_{\mu_1}(x_1) \ldots A_{\mu_n}(x_n) \Big)  & \nonumber\\
    +& \frac{1}{2!} \sum_{n,m=2}^\infty \frac{1}{n m} \int_{x_1 \ldots x_n} \int_{y_1 \ldots y_m} S_{\mu_1 \ldots \mu_n , \nu_1 \ldots \nu_m}(x_1 \ldots x_n ; y_1 \ldots y_m) \nonumber\\ 
    &&\hidewidth \text{tr} \Big( A_{\mu_1}(x_1) \ldots A_{\mu_n}(x_n) \Big) \text{tr} \Big( A_{\nu_1}(y_1) \ldots A_{\nu_m}(y_m) \Big) \nonumber\\
    +& \cdots \,.
\end{flalign}
From this we can see that single trace vertices are defined as cyclically symmetric in all indices,
\be
    S_{\mu_1 \ldots \mu_n}(x_1 \ldots x_n) = S_{\mu_n  \mu_1\ldots\mu_{n-1}}(x_n, x_1 \ldots x_{n-1}) \,,
\ee
whereas the remaining vertices are separately cyclically symmetric for each string.
However, the latter have been included for completeness only and are of no interest to us in the present paper. Therefore, from now on we will consider only the single trace vertices. The fact that the action $S$ is real, together with the hermiticity of the generators $T^a$, implies that $S^*_{\mu_1 \ldots \mu_n}(x_1 \ldots x_n) = S_{\mu_n \ldots \mu_1}(x_n \ldots x_1)$. This is an expression of charge conjugation invariance. To exploit translation invariance we work in momentum space where it implies momentum conservation:
\be
\label{Momentum space S}
    S_{\mu_1 \ldots \mu_n}(p_1 \ldots p_n) \, \delta \Big( \sum_i p_i \Big) = \int_{x_1 \ldots x_n} S_{\mu_1 \ldots \mu_n}(x_1 \ldots x_n) \: e^{- i \sum_i p_i \cdot x_i} \,.
\ee
By convention, we take all momenta pointing in towards the vertex. Note that the $S_{\mu_1 \ldots \mu_n}(p_1 \ldots p_n)$ are well-defined only when momentum is conserved. For the two-point vertex we write more simply $S_{\mu\nu}(p)=S_{\mu\nu}(p,-p)$. Lorentz invariance implies that:
\be
    S_{\mu_1 \ldots \mu_n}(p_1,\ldots, p_n) = (-1)^n S_{\mu_1 \ldots \mu_n}(-p_1,\ldots,-p_n) \,.
\ee
In other words, vertices with an even/odd number of legs are even/odd under a change of sign in all its momentum arguments.
Combined with charge conjugation invariance we thus also have that they are even/odd under reversal of their arguments
\be
    S_{\mu_n \ldots \mu_1}(p_n,\ldots, p_1) = (-1)^n S_{\mu_1 \ldots \mu_n}(p_1,\ldots,p_n) \,.
\ee
As an example we note that these symmetries tell us that a three-point action vertex is totally antisymmetric in its arguments $({p_1}_{\mu_1}, {p_2}_{\mu_2}, {p_3}_{\mu_3})$.

The flow equation \eqref{flowequation} manifestly preserves gauge invariance, namely \eqref{gaugetransf}, and this allows us to write a set of `trivial' Ward identities for the effective action $S$ that relates the longitudinal part of $n$-point vertices to differences of $(n-1)$-point vertices:
\be
\label{WardAction}
    {p_1}_{\mu_1} S_{\mu_1 \ldots \mu_n}(p_1 \ldots p_n) = S_{\mu_2 \ldots \mu_n}(p_1 + p_2, p_3 \ldots p_n) - S_{\mu_2 \ldots \mu_n}(p_2 \ldots p_{n-1}, p_1 + p_n) \,.
\ee
Since there are no $(n<2)$-point vertices, the two-point vertex is transverse:
\be 
p_\mu S_{\mu\nu}(p)= 0\,.
\ee
In the limit of small contracted momentum $p_1$, \eqref{WardAction} yields differential Ward identities \eg 
\be
\label{S3ptDiffWard}
    S_{\mu\nu\lambda}(0,p,-p) = \partial^p_\mu S_{\nu\lambda}(p)  \,.
\ee
where we have introduced the shorthand $\partial^p_\mu = \frac{\partial}{\partial p_\mu}$. This follows straightforwardly starting from \eqref{WardAction} after Taylor expanding its right-hand side:
\beal
    \epsilon_\mu S_{\mu\nu\lambda}(\epsilon,p,-p-\epsilon) &= S_{\nu\lambda}(p+\epsilon) - S_{\nu\lambda}(p) \\
    &= \epsilon_\mu \partial^p_\mu S_{\nu\lambda}(p) + O(\epsilon^2) \,, 
\eeal
recovering \eqref{S3ptDiffWard} from $O(\epsilon)$. We can proceed similarly about the four-point vertex to get:
\be
\label{S4ptDiffWard}
    S_{\mu\nu\lambda\sigma}(0,0,p,-p) + S_{\nu\mu\lambda\sigma}(0,0,p,-p) = \partial^p_\mu \partial^p_\nu S_{\lambda\sigma}(p)  \,.
\ee
The $I^{(n)}$ vertices already introduced in eqs. \eqref{I2}--\eqref{I4} satisfy the above properties.

\subsubsection{Kernel vertex properties}
\label{sec:Kernel vertices}

The regularisation and the flow equation are built up from constructs where actions are joined together by kernels to form new compound actions. Generically for two actions, say $R[A]$ and $S[A]$, the construct takes the form $R_{,a} K^{ab} S_{,b}$ and involves some kernel $K^{ab}[A]$. Since everything is constructed gauge covariantly, these kernels are themselves necessarily functionals of $A_\mu$. Although the kernels we need are constructed in a specific way, as reviewed in sec. \ref{sec:Regularisation structure}, they are special cases of a more general form of kernel namely any such that provides a \textit{covariantization} of some momentum kernel $K^{ab}[A=0]\equiv K_{\mu\nu}(p)$, \cf \eqref{K}, via gauge fields $A_\mu$ that act by commutation. (The latter property is assured here to the $2$-point level that we will require them, in essence by constructing the kernels using covariant derivatives $D_\mu$, see sec. \ref{sec:Higher-point vertices} and app. \ref{appendix:interleave identities}.) Such kernels have vertices whose properties generalise those for an action and ensure that the vertices of the compound action continue to satisfy the correct properties.

Expressions involving such a covariantized kernel $K_{\mu\nu}$ can be equivalently written in the following way:
\beal
    R_{,a} K^{ab} S_{,b} \equiv& \frac{1}{\Lambda^2} \frac{\delta R}{\delta A_\mu} \cdot K_{\mu\nu} \cdot \frac{\delta S}{\delta A_\nu} \\
    \equiv& \frac{1}{2\Lambda^2} \text{tr} \int_x \int_y \frac{\delta R}{\delta A_\mu(x)} K_{\mu\nu}(x,y) \cdot \frac{\delta S}{\delta A_\nu(y)} \,,
\eeal
where we suppress the functional dependence on the gauge field.
Taylor expanding with respect to the gauge field in $K^{ab}[A]$ gives a series expansion in nested commutators \cite{Morris:1999px,Morris:2000fs,Arnone:2002cs}:
\besp
\label{KernelCommutatorsRep}
    R_{,a} {K}^{ab} S_{,b} \equiv \frac{1}{2\Lambda^2}\sum_{n=0} \int_x \int_y \int_{x_1 \cdots x_n} K_{\mu_1 \ldots \mu_n;\mu\nu}(x_1 \ldots x_n; x, y) \\
    \text{tr} \Big[ \frac{\delta R}{\delta A_\mu(x)} A_{\mu_1}(x_1) \cdots A_{\mu_n}(x_n) \cdot \frac{\delta S}{\delta A_\nu(y)} \Big] \,.
\eesp
Now, expanding the commutators, we obtain:
\besp
\label{KernelExplicitRep}
    R_{,a} {K}^{ab} S_{,b} \equiv \frac{1}{2\Lambda^2}\sum_{m,n=0} \int_x \int_y \int_{x_1 \cdots x_n , y_1 \cdots y_m} K_{\mu_1 \ldots \mu_n , \nu_1 \ldots \nu_m ;\mu\nu}(x_1 \ldots x_n; y_1 \ldots y_m ; x, y) \\
    \text{tr} \Bigg[ \frac{\delta R}{\delta A_\mu(x)} A_{\mu_1}(x_1) \cdots A_{\mu_n}(x_n) \frac{\delta S}{\delta A_\nu(y)} A_{\nu_1}(y_1) \cdots A_{\nu_m}(y_m) \Bigg] \,.
\eesp
In momentum space, kernel vertices take the following form:
\besp
    K_{\mu_1 \ldots \mu_n , \nu_1 \ldots \nu_m ;\mu\nu}(p_1 \ldots p_n; q_1 \ldots q_m ; p, q) \, \delta \Big( \sum p_i + \sum q_i + p + q \Big) = \\
    = \int_x \int_y \int_{x_1 \cdots x_n , y_1 \cdots y_m} K_{\mu_1 \ldots \mu_n , \nu_1 \ldots \nu_m ;\mu\nu}(x_1 \ldots x_n; y_1 \ldots y_m ; x, y) e^{-i \Big( p \cdot x + q \cdot y + \sum_i p_i \cdot x_i + \sum_j q_j \cdot y_j \Big)} \,.
\eesp
The two sets of vertices in \eqref{KernelCommutatorsRep} and \eqref{KernelExplicitRep} are related through `interleave' identities (also called `coincident line' identities \cite{Morris:1999px,Morris:2000fs,Arnone:2002cs}) which here just express the fact that covariantisation is via commutation:
\be
\label{interleave identities}
    K_{\mu_1 \ldots \mu_n , \nu_1 \ldots \nu_m ;\mu\nu}(p_1 \ldots p_n; q_1 \ldots q_m ; p, q) = (-1)^m\!\!\! \sum_\text{interleaves} K_{\alpha_1 \ldots \alpha_{n+m};\mu\nu}(k_1 \ldots k_{m+n}; p , q) \,.
\ee
In the above the sum runs over all the possible arrangements of the combined sequence of $p_1^{\mu_1} \ldots p_n^{\mu_n}$ and $q_1^{\nu_1} \ldots q_m^{\nu_m}$, in which the $p$ momenta remain ordered with one another, whereas the $q$ momenta order is reversed (via the so-called shuffle product $\sum k^{\alpha_1}_1\ldots k^{\alpha_{m+n}}_{m+n} = p_1^{\mu_1} \ldots p_n^{\mu_n}\shuffle q_m^{\nu_m} \ldots q_1^{\nu_1}$).

The case where $m=n=0$ (thus leaving a double semi-colon) is of course just the momentum kernel again:
\be
    K_{ , ;\mu\nu}( ; ;p,-p) = K_{\mu\nu}(p) \,.
\ee
We write the $m=0$ case more simply as:
\be
    K_{\mu_1 \ldots \mu_n , ;\mu\nu}(p_1 \ldots p_n; ; p, q) = K_{\mu_1 \ldots \mu_n;\mu\nu}(p_1 \ldots p_n; p, q) \,,
\ee
then from \eqref{interleave identities} the $n=0$ case is given by
\be
    K_{ , \nu_1 \ldots \nu_m ;\mu\nu}( ; q_1 \ldots q_m ; p, q) = (-1)^m K_{\nu_m \ldots \nu_1;\mu\nu}(q_m \ldots q_1 ; p, q) \,.
\ee
A useful special case of \eqref{interleave identities} is:
\be 
  K_{\mu,\nu;\alpha\beta}(p;q;r,s) = -K_{\mu\nu;\alpha\beta}(p,q;r,s)-K_{\nu\mu;\alpha\beta}(q,p;r,s)\,.
\ee

Kernel vertices obey Ward identities closely similar to \eqref{WardAction} \cite{Morris:1999px,Morris:2000fs}. Furthermore the contracted momentum can reach the end of the string of momenta, where it attaches to the outer momentum:
\besp
\label{WardKernel}
    {p_1}_{\mu_1} K_{\mu_1 \ldots \mu_n , \nu_1 \ldots \nu_m ; \mu \nu}(p_1 \ldots p_n ; q_1 \ldots q_m ; p , q) = \\ 
    = K_{\mu_2 \ldots \mu_n , \nu_1 \ldots \nu_m ; \mu \nu}(p_1 + p_2, p_3 \ldots p_n ; q_1 \ldots q_m ; p , q) \\
    - K_{\mu_2 \ldots \mu_n , \nu_1 \ldots \nu_m ; \mu \nu}(p_2 \ldots p_n ; q_1 \ldots q_m ; p_1 + p , q) \,.
\eesp
Again one can derive differential Ward identities (\cf \eqref{S3ptDiffWard}, \eqref{S4ptDiffWard}). We list below some of the more useful ones:
\beal
    K_{\lambda;\mu\nu}(0;-p,p) &= \partial^p_\lambda K_{\mu\nu}(p) \label{1ptDiffWardKernel} \,, \\
    K_{\lambda\sigma;\mu\nu}(-p,0;0,p) &= \partial^p_\sigma K_{\lambda;\mu\nu}(-p;0,p) \label{2ptDiffWardKernel} \,.
\eeal

\subsubsection{Two-point action vertices}
\label{sec:twopoints}

Combining these definitions with those of sec. \ref{sec:Background independent regularisation} one can derive the expressions for the vertices that are needed and work through the classical and quantum corrections systematically. Thus for example, from the definition \eqref{A} of the quasi-local  action $A$, combined with that of $X$, \eqref{X}, and the fundamental two-point vertex $I^{(2)}$ in \eqref{I2}, or alternatively directly from \eqref{2ptA}, one gets its two-point vertex:
\be
    A_{\mu_1\mu_2}(p) = 2 \, \square_{\mu_1\mu_2}(p) F_A(p) \,,\qquad
    F_A(p) := \frac{1}{c_p} \,.
\ee
Armed with this one finds from \eqref{seed action} the two-point vertex for 
the seed action:
\be
\label{Shattwopt}
    \hat{S}_{\mu_1\mu_2}(p) = 2 \, \square_{\mu_1\mu_2}(p) F_{\hat{S}}(p) \,, \qquad
    F_{\hat{S}}(p) := \frac{\Tilde{p}^2 + c_p}{c_p^2} \,,
\ee
and from \eqref{S0} the classical effective action $S^0$ two-point vertex:
\be
\label{S0twopt}
    S^0_{\mu_1\mu_2}(p) = 2 \, \square_{\mu_1\mu_2}(p) F(p) \,, \qquad
    F(p) := 1 + \frac{\Tilde{p}^2}{c_p^2} \,.
\ee
From the definition \eqref{Sigma} of $\Sigma$ and $X$ one gets immediately its classical two-point vertex:
\be
\label{Sigma2pt}
    \Sigma^0_{\mu_1\mu_2}(p) = 2 \, \square_{\mu_1\mu_2}(p) F_\Sigma(p) \,, \qquad
    F_\Sigma(p) := 1 - \frac{1}{c_p} \,.
\ee
Finally from its definition \eqref{theta} we get the classical two-point vertex for the action $\theta$:
\be
    \theta^0_{\mu_1\mu_2}(p) = 2 \, \square_{\mu_1\mu_2}(p) F_\theta(p)\,,\qquad
    F_\theta(p) := \frac{c_p - 2 \tilde{p}^2 c_p'}{c_p^2} \,.
\ee

\subsubsection{Zero-point kernel functions}
\label{sec:zeropoints}

The zero-point vertices for the \textit{simple} (\ie non-compound) kernels $X$, $Y$, $B$, $b$, $t$, $\kappa$ and $W$, follow straightforwardly from their definitions in \eqref{X}, \eqref{B}, \eqref{b}, \eqref{kappakernel} and \eqref{theta}, by replacing their argument ($\Delta$ or $\Delta_\parallel$ as appropriate) by $\tilde{p}^2$, for example:
\be
    \kappa(p) = 2 \, \frac{c_p - 2 \Tilde{p}^2 c^{\prime}_p}{c_p^{2} + \Tilde{p}^2} \,.
\ee
(The zero-point $X$ kernel was in effect already used to derive the two-point action vertices above.)
Those for the \textit{compound} kernels follow almost as straightforwardly. Thus the zero point vertex for $\big( C^{-1} \big)^a{}_b$ follows from its definition \eqref{C^-1} since the generator of gauge transformations \eqref{generators} collapses to a partial derivative in this case. Writing 
\be
\label{Cm0pt}
    C^{-1}_{\mu\nu}(p) = \frac{1}{C_T(p)} \frac{\square_{\mu\nu}(p)}{p^2}  + \frac{1}{C_L(p)} \frac{p_\mu p_\nu}{p^2} \,,
\ee
we have that its longitudinal and transverse functions are given by:
\be
    C_T(p) := \frac{c_p}{c_p + \Tilde{p}^2} \,, \qquad
    C_L(p) := \frac{1}{1 + \Tilde{p}^2 Y_p} \,.
\ee
Note that despite appearances in \eqref{Cm0pt}, the kernel has a Taylor expansion in $p_\mu$ (is quasilocal) because $Y$ is quasilocal and $c_p$ is normalised to $c(0)=1$ (\cf below \eqref{X}). Inverting gives us the zero-point vertex for $C^a{}_b$:
\be
\label{C}
    C_{\mu\nu}(p) = C_T(p) \frac{\square_{\mu\nu}(p)}{p^2}  + C_L(p) \frac{p_\mu p_\nu}{p^2} \,,
\ee
which is also quasi-local for the same reasons. Then that for $\mathcal{K}$ follows from \eqref{Kkernel}:
\be
    \mathcal{K}_{\mu\nu}(p) = \mathcal{K}_T(p) \frac{\square_{\mu\nu}(p)}{p^2}  + \mathcal{K}_L(p) \frac{p_\mu p_\nu}{p^2} \,,\qquad \mathcal{K}_I(p) := \kappa(p) C_I(p)\quad (I=T,L)\,.
\ee

\subsubsection{Higher-point vertices}
\label{sec:Higher-point vertices}

The $n$-point vertices for the simple kernels can be computed by Taylor expanding in $\Delta$ or $\Delta_\parallel$ as appropriate. Keeping only $n$ instances of $A_{\mu_i}$ as in \eqref{KernelCommutatorsRep}, the momentum dependence can then be resummed to give the explicit formula. The vertices for those that are simply functions of $\Delta_\parallel=-D^2/\Lambda^2$ -- namely $Y$, $B$, $b$ and $t$ -- have vertices that have already been computed this way in sec. 5 of ref. \cite{Morris:2000fs}. We just quote the results:
\beal Y_\mu(p;r,s) &=(r-s)_\mu\,\frac{Y_r-Y_s}{ p\!\cdot\!(r-s)}\,,\label{Y1pt} \\
Y_{\mu\nu}(p,q;r,s) &=\delta_{\mu\nu}\frac{Y_s-Y_r}{s^2-r^2}-(p+2r)_\mu(q+2s)_\nu\,\Xi_Y(p,q;r,s)\,,\\
\Xi_Y(p,q;r,s) &=\frac{Y_{s+q}}{q\!\cdot\!(q+2s)\,p\!\cdot\!(p+2r)}
+\frac{1}{ s^2-r^2}\left[\frac{Y_r}{ p\!\cdot\!(p+2r)}-\frac{Y_s}{ q\!\cdot\!(q+2s)}\right]\,,
\eeal
where we have introduced the shorthand $\Xi_Y$, and where for the others one simply replaces $Y$ by $B$, $b$ or $t$ as appropriate. We note that these equations need care at special momenta where  denominators vanish. For example the correct equation for $Y_{\mu\nu}(p,-p;r,-r)$ follows from $\lim_{\epsilon\to0}Y_{\mu\nu}(p,-p-\epsilon;r+\epsilon,-r)$. Again we refer to sec. 5 of ref. \cite{Morris:2000fs} for details. The vertices for the other simple kernels $X$, $\kappa$ and $W$ follow from using a similar strategy. We only have to take into account the presence of $F_{\mu\nu}$ when expanding in a series of powers of $\Delta_{\mu\nu}$, \cf \eqref{Delta}. Thus we find: 
\beal 
\label{X1pt}
X_{\lambda;\mu\nu}(p;r,s) &= \left\{ (r-s)_\lambda\delta_{\mu\nu}+4\delta_{\lambda[\mu}p_{\nu]} 
\right\} \frac{X_r-X_s}{ p\!\cdot\!(r-s)}\,,\\
X_{\lambda\sigma;\mu\nu}(p,q;r,s) &=\left(\delta_{\mu\nu}\delta_{\lambda\sigma}+4\delta_{\lambda[\mu}\delta_{\nu]\sigma}\right)\frac{X_s-X_r}{s^2-r^2}+\Bigg(16\delta_{\lambda[\mu}p_{\epsilon]}\delta_{\sigma[\epsilon}q_{\nu]}-4\delta_{\lambda[\mu}p_{\nu]}(q+2s)_\sigma\nn\\
&\quad +4(p+2r)_\lambda\delta_{\sigma[\mu}q_{\nu]}-\delta_{\mu\nu}(p+2r)_\lambda(q+2s)_\sigma\Bigg) \Xi_X(p,q;r,s)\,.
\eeal
Here $T_{[\mu\nu]} := \frac12( T_{\mu\nu}-T_{\nu\mu})$. As above, those for $\kappa$ and $W$ follow simply by replacing the name. 

With the above building blocks and techniques it is a tedious but straightforward exercise to arrive at explicit expressions for compound kernel and action $n$-point vertices. For example from \eqref{A}, one finds
\beal
\label{A3pt}
&A_{\mu\nu\lambda}(p,q,-p-q) = I_{\mu\nu\lambda}(p,q,-p-q)+\frac{1}{2\Lambda^2}\Bigg[I_{\alpha\mu\nu}(-p-q,p,q)X_{p+q}I_{\alpha\lambda}(p+q)\nn\\
&+I_{\alpha\nu\lambda}(p,q,-p-q)X_pI_{\alpha\mu}(p)+I_{\alpha\lambda\mu}(q,-p-q,p)X_qI_{\alpha\nu}(q)+I_{\alpha\mu}(p)X_{\nu;\alpha\beta}(q;p,-p-q)I_{\beta\lambda}(p+q)\nn\\
&+I_{\alpha\nu}(q)X_{\lambda;\alpha\beta}(-p-q;q,p)I_{\beta\mu}(p)+I_{\alpha\lambda}(p+q)X_{\mu;\alpha\beta}(p;-p-q,q)I_{\beta\nu}(q)\Bigg]\,.
\eeal
Using \eqref{C^-1}, this is sufficient to construct the one-point vertex for the compound kernel $\big( C^{-1} \big)^a{}_b$: 
\be 
\label{Cm1pt}
C^{-1}_{\mu;\alpha\beta}(p;r,s) =\frac1{2\Lambda^2}A_{\mu\beta\alpha}(p,s,r)+\frac1{\Lambda^2}\left\{\delta_{\mu\alpha}s_\beta Y_s-r_\alpha\delta_{\mu\beta}Y_r-r_\alpha s_\beta Y_\mu(p;r,s)\right\}\,,
\ee
and from this we also have the $C$ one-point vertex:
\be 
\label{C1pt}
C_{\mu;\alpha\beta}(p;r,s) = - C_{\alpha\gamma}(r) C^{-1}_{\mu;\gamma\delta}(p;r,s) C_{\delta\beta}(s) \,.
\ee
We give further details on the construction of the compound kernels in app. \ref{appendix:interleave identities} where we also prove that up to the two-point level the compound kernels obey the interleave identities \eqref{interleave identities}.
Note that, as a consequence of their symmetric definition as operators, and the interleave identities, all the kernel one-point vertices so far treated are antisymmetric under $r_\alpha \leftrightarrow s_\beta$, for example:
\be\label{antisymm1pt} C^{-1}_{\mu;\alpha\beta}(p;r,s) = C^{-1}_{,\mu;\beta\alpha}(;p;s,r) = -C^{-1}_{\mu;\beta\alpha}(p;s,r)\,. \ee
The definition \eqref{Kkernel}, $\K^{ab}=C^a{}_c\kappa^{cb}$, is not symmetric, and thus its one-point vertex does not obey this identity.
However, using \eqref{C1pt}, and \eqref{X1pt} with $X\mapsto\kappa$, we get an explicit expression for it also:
\be\label{K1pt} \rK_{\mu;\alpha\beta}(p;r,s) = C_{\mu;\alpha\beta}(p;r,s)\,\kappa(s)+C_{\alpha\gamma}(r)\,\kappa_{\mu;\gamma\beta}(p;r,s)\,. \ee
We should distinguish this from the vertex made from the hermitian conjugate $\kappa^{ac}C_c{}^b$:
\be\label{K1ptb} \lK_{\!\mu;\alpha\beta}(p;r,s) = \kappa(r)\, C_{\mu;\alpha\beta}(p;r,s)+\kappa_{\mu;\alpha\gamma}(p;r,s)\,C_{\gamma\beta}(s)\,, \ee
and the symmetrised combination 
\be \lrK_{\!\mu;\alpha\beta}(p;r,s)=\frac12\left[ \rK_{\mu;\alpha\beta}(p;r,s) + \lK_{\mu;\alpha\beta}(p;r,s) \right]\,, \ee
coming from $\frac12( C^a{}_c\kappa^{cb} + \kappa^{ac}C_c{}^b )$, since these versions can also appear.
As a consequence of the interleave identities, $\lrK_{\mu;\alpha\beta}(p;r,s)$ obeys \eqref{antisymm1pt}, whilst the directed versions satisfy
\be \rK_{\mu;\alpha\beta}(p;r,s) = -\lK_{\!\mu;\beta\alpha}(p;s,r)\,. \ee

\subsubsection{Large momentum behaviour}
\label{sec:Large momentum behaviour}

As we will see in sec. \ref{sec:One loop beta function}, these vertices play a closely similar r\^ole to those of Feynman rules. In particular in one-loop diagrams, three-point vertices will carry a loop momentum $q$ and external momentum $p$ and thus have arguments $p$, $q$, and $-p-q$. Therefore the large momentum behaviour that actually determines whether the flow equation is properly regularised is one where $q\to\infty$ while keeping $p$ fixed. As before, it is sufficient to characterise the large momentum behaviour in terms of the power of $c_q$, since the proposed regularisation structure works only if the quantum corrections are regularised overall by some negative power of $c_q$.

In fact the three-point action vertices\footnote{Recall that three-point action vertices are totally antisymmetric so the order of the arguments is irrelevant.} $S_{\mu\nu\lambda}(p,q,-p-q)$ diverge at least as rapidly in terms of $c_q$, as their corresponding two-point action vertices $S_{\mu\lambda}(q)$. This follows from the Ward identity:
\be 
\label{S3ptWard}
q^\nu S_{\mu\nu\lambda}(p,q,-p-q) = S_{\mu\lambda}(p) - S_{\mu\lambda}(p+q)\,. \ee
Similarly the degree of divergence of a one-point kernel is set by the zero-point kernel. If $q$ goes from end to end, the one-point kernel  behaves with at least the same power of $c_q$ as $K_q$:
\be p^\lambda K_{\lambda;\mu\nu}(p;q,-p-q) = K_{\mu\nu}(q) -K_{\mu\nu}(q+p)\,.\ee
But if $q$ passes through the side, the large $q$ behaviour depends on whether $K_q$ diverges or decays:
\be q^\lambda  K_{\lambda;\mu\nu}(q;p,-p-q) = K_{\mu\nu}(p) -K_{\mu\nu}(q+p)\,.\ee
If $K_q$ diverges, then the one-point kernel diverges at least as rapidly, whilst if $K_q$ decays then $K_{\lambda;\mu\nu}(q;p,-p-q)$ cannot decay, because the right hand side of the Ward identity has  $K_{\mu\nu}(p)$ which is independent of $q$.

However, Ward identities only set a lower bound on the power of $c_q$, because there remains the possibility that vertices have a worse behaving transverse part. 
We now pin down the precise behaviour as a power of $c_q$, \ie both longitudinal and transverse parts.\footnote{For a detailed analysis of the large momentum behaviour of arbitrary $n$ higher-point (simple) kernel vertices see sec. 5 of ref. \cite{Morris:2000fs}.}

We start by analysing $A^{(3)}$. This is displayed in \eqref{A3pt}. The fundamental vertices $I^{(n)}$ in \eqref{A3pt} only provide some power of $q$ at worst, \cf \eqref{I2} -- \eqref{I4}, so $A^{(3)}$s large $q$ behaviour is set by its kernels. From \eqref{X} we have $X_q \approx X_{p+q} \approx c_q^{-1}$, and from \eqref{X1pt} we see that all the  $X$ one-point vertices in \eqref{A3pt} also diverge in this way. Two terms in $A^{(3)}$ do not diverge with the cutoff, namely the pure $I$ term (\ie the first term) and the term containing $X_p$ on the second line, but the others dominate, so overall $A^{(3)} \approx c^{-1}_q$, as previously assumed, and matching its two-point vertex \eqref{2ptA} as per the minimum required by the Ward identity.

Since from \eqref{theta0} the classical action $\theta_0$ has the same formula as \eqref{A3pt}, only with the replacement $X\mapsto W$, we see immediately that its three-point vertex also $\approx c^{-1}_q$. In fact, given that the quantum correction is $\propto A$, this holds for the full $\theta^{(3)}$.

We can borrow the same formula also for the seed action $\hat{S}$. From \eqref{seed action} we replace $I\mapsto A$ and $X\mapsto 1$ on the right hand side of \eqref{A3pt} (and so delete the three terms involving the $X$ one-point vertex). Since we have already established that $A_{\mu\nu\lambda}(p,q,-p-q)\approx c^{-1}_q$, we see that indeed $\hat{S}_{\mu\nu\lambda}(p,q,-p-q)\approx c^{-2}_q$ again as previously assumed and enforced by the Ward identity.

It is not hard to see that these estimates hold also for the higher point $A^{(n)}$, $\theta^{(n)}$ and $\hat{S}^{(n)}$ action vertices, 
where again two of their momentum arguments diverge as $\sim\pm q$ with the others held fixed. For the four-point example, their explicit formulae can be read off from the right hand side of \eqref{4pt classical S0}. The same is true for higher point kernel vertices where the zero-point kernel diverges at large momentum. However we will only use the estimates for the three-point vertices in this paper.

At one loop we need also the classical effective action three-point vertex and this in turn depends on the kernels $\K$ and $C$ and their one-point vertices (see eqn. \eqref{3pt classical S0 flow equation}). From \eqref{C} we clearly have $C_{\alpha\beta}(q)\approx c_q$ (recall $Y_q\approx c^{-1}_q$). 
Clearly from \eqref{Cm1pt}, the $C^{-1}$ one-point vertex $\approx c^{-1}_q$ whatever the arrangement of the arguments $p$, $q$, and $-p-q$, this being the minimum imposed by the Ward identity. But from  \eqref{C1pt} the arrangement of the arguments matters for $C$:
\beal
    C_{\mu;\alpha\beta}(p;q,-p-q) \approx C_q &\approx c_q \,,\label{endtoend} \\
    C_{\nu;\alpha\beta}(q;p,-p-q) &\approx 1 \,,
\eeal
again precisely as predicted by the Ward identities.
One can check using the explicit formulae from the previous section that the behaviour predicted by the Ward identities, holds for all the kernel vertices whose zero-point kernels decay at large momentum, namely $t$, $\kappa$, $C$ and $\K$. 
In particular we have also 
\beal
    \K_{\mu;\alpha\beta}(p;q,-p-q) \approx \K_q &\approx c^2_q \,, \label{Kendtoend}\\
    \K_{\nu;\alpha\beta}(q;p,-p-q) &\approx 1 \,.
\eeal

\section{Regularisation of higher point classical vertices}
\label{sec:Regularisation of higher point classical vertices}

As we saw in sec. \ref{sec:Background independent regularisation}, the space-time trace terms are designed to implement PV regularisation of the $S_{UV} \approx c^{-2}_q$ part. We will see in sec. \ref{sec:secondorder} that there are subtleties with this however.  The remaining regularisation works provided $\mathcal{S}^{(n\ge3)}_0$ corrections diverge slower than $c^{-2}_q$. This was the key property set out in eqns. \eqref{S0} and \eqref{assumption1}, where it was assumed that in fact $\mathcal{S}^{(n)}_0\approx c^{-1}_q$, or equivalently $\Sigma^{(n)}_0\approx c^{-1}_q$. For the three-point we can now state this more precisely as:
\be 
\label{assumption3}
\Sigma^0_{\beta\nu\alpha}(q+p, -p, -q)\approx c^{-1}_q\,. 
\ee 
In this section we confirm in detail that this estimate holds true. It is then clear that the estimates are correct for all the higher point $\Sigma^{(n)}_0$ vertices also.

In order to compute $\Sigma^{(3)}_0$ we need the classical action three-point vertex. The latter follows from the flow equation $\Lambda D_\Lambda S^0=0$. The analysis is simpler if instead we recast it directly as a flow for $\Sigma_0$.
Taking the classical limit (\ie $O(1/g^2)$ in the loop expansion) of \eqref{flowequation2}, and using the identity \eqref{Shatid}, we have
\be
\label{classicalSigma}
     \dot{\Sigma}^0 = - \dot{A} + \frac{1}{2} \Sigma^0_{,a} \K^{ab} \Sigma^0_{,b} + \frac{1}{2} {A}_{,a} \kappa^{ab} \Sigma^0_{,b} - \Sigma^0_{,a} C^{ab} \theta^0_{,b} \,.
\ee
Expanding to the three-point level:
\beal
\label{classicalSigma3pt}
\begin{split}
    \dot{\Sigma}^0_{\mu\nu\lambda}(p,q,-p-q) = &- \dot{A}^0_{\mu\nu\lambda}(p,q,-p-q) \\
    &+ \frac{1}{2\Lambda^2} \Sigma^0_{\mu\alpha}(p) \mathcal{K}_T(p) \Sigma^0_{\alpha\nu\lambda}(p,q,-p-q) \\
    &+ \frac{1}{2\Lambda^2}\Sigma^0_{\mu\alpha}(p) \lrK_{\!\nu;\alpha\beta}(q;p,-p-q)  \Sigma^0_{\beta\lambda}(p+q) \\
    &+ \frac{1}{4\Lambda^2} \Sigma^0_{\alpha\mu}(p) \kappa(p) A_{\alpha\nu\lambda}(p,q,-p-q)  \\
    &+ \frac{1}{4\Lambda^2}\Sigma^0_{\mu\nu\alpha}(p,q,-p-q) \kappa(p+q) A_{\alpha\lambda}(p+q) \\
    &+ \frac{1}{4\Lambda^2}\Sigma^0_{\mu\alpha}(p) \kappa_{\nu;\alpha\beta}(q;p,-p-q) A_{\beta\lambda}(p+q) \\
    &- \frac{1}{4\Lambda^2}\Sigma^0_{\mu\alpha}(p) \kappa_{\lambda;\alpha\beta}(-p-q;p,q) A_{\beta\nu}(q) \\
    &- \frac{1}{2\Lambda^2} \Sigma^0_{\alpha\mu}(p) C_T(p) \theta^0_{\alpha\nu\lambda}(p,q,-p-q)  \\
    &- \frac{1}{2\Lambda^2}\Sigma^0_{\mu\nu\alpha}(p,q,-p-q) C_T(p+q) \theta^0_{\beta\lambda}(p+q) \\
    &- \frac{1}{2\Lambda^2}\Sigma^0_{\mu\alpha}(p) C_{\nu;\alpha\beta}(q;p,-p-q) \theta^0_{\beta\lambda}(p+q) \\
    &+ \frac{1}{2\Lambda^2}\Sigma^0_{\mu\alpha}(p) C_{\lambda;\alpha\beta}(-p-q;p,q) \theta^0_{\beta\nu}(q) \\
    &+ \text{cycles} \,.
\end{split}
\eeal
Here we have used properties of the kernels. First of all, if we contract a zero-point kernel into a two-point vertex, \eg as on second line, then only its transverse part contributes because of the transversality of the two-point function. In addition to this, we used that $\K^a{}_b$, $C^a{}_b$ and $\kappa^a{}_b$, kernel one-point vertices obey interleave identities \eqref{interleave identities}. 

Those terms that depend on $\Sigma_0^{(3)}$ in \eqref{classicalSigma3pt} we take to the left of the equation. Their multiplying factors collect into 
\be \label{Z} Z_{\mu\nu}(p_i)=\K_T(p_i)\, \Sigma^0_{\mu\nu}(p_i) \ee 
for each external momentum $p_i$, as follows from the identity
\be 
\label{Cthetaidentity}
C_T(p)\,\theta^0_{\mu\nu}(p) = \frac12 \K_T(p)\, S^0_{\mu\nu}(p)
\ee
(this can be shown from their explicit form given in  secs. \ref{sec:twopoints}, \ref{sec:zeropoints}) and $\kappa(p) A_{\mu\nu}(p) = \K_T(p)\hat{S}_{\mu\nu}(p)$, which is the identity \eqref{Shatid} at two-point level. Using the results of the previous section it is straightforward to verify explicitly that the remaining terms, which we call $T_{\mu\nu\lambda}(p,q,-p-q)$, diverge as $T_{\mu\nu\lambda}(p,q,-p-q)\approx c^{-1}_q$. Indeed, given that in \eqref{classicalSigma} the action vertices diverge as $c^{-1}_q$, the only way to produce a contribution that diverges faster than this would be to have the high momentum flow, $q$, go through both action factors. But this requires $q$ to flow through the kernel from end to end, thus reducing the divergence by at least a factor of $c_q$ again ($c_q$ for $\kappa$ and $C$, while $\K$ would provide $c^2_q$). In this way we verify $\Sigma^{(n)}_0 \approx c_q^{-1}$ for all $n$, \ie \eqref{assumption2} at the classical level.

The flow at the three-point level can be written now as
\besp
\label{failurevertex}
\Lambda\partial_\Lambda \Sigma^0_{\mu\nu\lambda}(p,q,-p-q)\\ - \frac1{4\Lambda^2}\Big[ Z_{\mu\alpha}(p) \Sigma^0_{\alpha\nu\lambda}(p,q,-p-q)
    + Z_{\nu\alpha}(q) \Sigma^0_{\alpha\lambda\mu}(q,-p-q,p)  + Z_{\lambda\alpha}(p+q) \Sigma^0_{\alpha\mu\nu}(-p-q,p,q) \Big]\\
    = T_{\mu\nu\lambda}(p,q,-p-q)\,.
\eesp
This is in a form where it can also be integrated with respect to $\Lambda$. $T_{\mu\nu\lambda}(p,q,-p-q)$ is a known function which can be constructed explicitly from the vertices discussed in the previous sections. The terms in square brackets on the left hand side of \eqref{failurevertex}  provide the integrating factor for the differential equation. Regarding $Z_{\mu\nu}(p)$ as a matrix $Z(p)$, we define
\be 
\label{zeta}
\zeta_{\mu\nu}(p) = \left[ \exp \int_\Lambda^\infty \frac{d\Lambda_1}{4\Lambda^3_1}\, Z(p) \right]_{\mu\nu}  = \delta_{\mu\nu} + \frac{\Box_{\mu\nu}(p)}{p^2}\left(-1+\exp \int_\Lambda^\infty \frac{d\Lambda_1}{2\Lambda_1}\, \tilde{p}^2\K_T(p)F_\Sigma(p)\right)\,,
\ee
where for the second equality we use \eqref{Z} and \eqref{Sigma2pt}.
The integrated $\Sigma^{(3)}_0$ is then given by
\besp 
\label{sigmaint}
\Sigma^0_{\mu\nu\lambda}(p,q,-p-q) =\\ - \zeta^{-1}_{\mu\alpha}(p)\,\zeta^{-1}_{\nu\beta}(q)\,\zeta^{-1}_{\lambda\gamma}(-p-q) \int^\infty_\Lambda \frac{d\Lambda_1}{\Lambda_1}\, \zeta_{\alpha\alpha'}(p)\,\zeta_{\beta\beta'}(q)\,\zeta_{\gamma\gamma'}(-p-q)\,T_{\alpha'\beta'\gamma'}(p,q,-p-q)\,,
\eesp
where $\zeta^{-1}_{\mu\nu}(p)$ is of course given by the same expression as \eqref{zeta} except for a minus sign in the exponential, and it is understood that all terms under a $\Lambda_1$-integral are evaluated at cutoff scale $\Lambda_1$. Note that the integration constant vanishes because, by gauge invariance and dimensions, $\Sigma_0$ vanishes in the limit $\Lambda\to\infty$.\footnote{As can be seen by Taylor expanding \eqref{Sigma2pt} for example, $\Sigma_0$ has a minimum of four space-time derivatives and thus has an overall $1/\Lambda^2$ factor.} Expanding the exponential in \eqref{zeta} we see that all the corrections that arise from these integrating factors are also transverse. In this sense the full vertex involves in fact an exponentiation of the equations of motion. It does not affect the divergence with $c_q$ however, since $Z(q)$ vanishes in the large $q$ limit.
The integrated version of the $n\ge4$-point vertices can be explicitly written down in a similar way.

\section{One loop beta function}
\label{sec:One loop beta function}

To see why it is the large $q$ behaviour of the above vertices that is important, and to provide a test of the formalism, we focus on the simplest quantum correction: the one-loop contribution to the effective action two-point vertex. Since gauge invariance is exactly preserved this computation should, if the flow equation is regularised correctly, also yield the one-loop beta function (i.e. $\beta_1 = \eta_1/2$).
To extract it, we need to define the coupling constant $g(\Lambda)$ beyond classical level. We do this by imposing a convenient \textit{renormalization condition}:
\be
\label{RenormalizationCondition}
    S = \frac{1}{2 g^2(\Lambda)} \text{tr} \int_x F_{\mu\nu}^2 + O(\partial^3) \,,
\ee
which means that the full two-point function of $S$ at $O(p^2)$ is given by:
\be
    S_{\mu_1\mu_2}(p) \big|_{O(p^2)} = 2 \: \square_{\mu_1\mu_2}(p) \:.
\ee
Extracting the $O(p^2)$ part of \eqref{S0twopt} we have that:
\be
    S^0_{\mu_1\mu_2}(p) \big|_{O(p^2)} = 2 \, \square_{\mu_1\mu_2}(p) = S_{\mu_1\mu_2}(p) \big|_{O(p^2)} \:.
\ee
This means that the renormalization condition at $O(p^2)$ is already saturated at tree level, and, thus, all higher order loop contributions must vanish,
\beal
    S^n_{\mu_1\mu_2}(p) \big|_{O(p^2)} = 0 &&, \forall n \geq 1 \,.
\eeal
Since $\Sigma^0 = S^0 - \hat{S}$ and $\Sigma^n = S^n$, we also obtain that:
\begin{align}
    \Sigma^n_{\mu_1\mu_2}(p) \big|_{O(p^2)} &= 0 & &, \: \forall n \geq 1 \:.
\end{align}
The remaining action $\theta$ behaves differently with respect to $S$ and $\Sigma$. Its loopwise expansion was already given in \eqref{thetan}
and from there we see that at $O(p^2)$ we have a non-vanishing higher loop contribution from its two-point vertex:
\be
    \theta^n_{\mu_1\mu_2}(p) \big|_{O(p^2)} = \eta_n \square_{\mu_1\mu_2}(p) .
\ee
We computed explicitly all the relevant classical action two-point vertex functions in sec. \ref{sec:twopoints}.

At this point we are able to recast \eqref{1loopB} as an algebraic equation for $\beta_1$. If we look at the flow equation of $S^1_{\mu\nu}(p)$ and restrict it to $O(p^2)$, we obtain:
\besp
\label{1loop at O(p^2)}
     -4 \beta_1 \square_{\mu\nu}(p) =  \text{Tr} \Bigg\{ \Lambda \partial_\Lambda \Big[ \ln \big( C^{-1} \big)^a{}_b \Big] - \Lambda \partial_\Lambda \big[ \ln B^\alpha{}_\beta \big] - \frac{1}{2} \Lambda \partial_\Lambda \big[ \ln b^\alpha{}_\beta \big] \\ 
     + \Psi_0^d \frac{\delta}{\delta \phi^d} \Big[ \ln \big( C^{-1} \big)^a{}_b \Big] - \Psi_0^a \frac{\delta}{\delta \phi^a} \big[ \ln B^\alpha{}_\beta \big] - \frac{1}{2} \Psi_0^a \frac{\delta}{\delta \phi^a} \big[ \ln b^\alpha{}_\beta \big] + \frac{\delta \Psi_0^a}{\delta \phi^b} \Bigg\}_{\substack{\text{2-pt.function} \\ \text{at }O(p^2)}} \,.
\eesp
Here we have used the requirement imposed on us by the renormalization condition \eqref{RenormalizationCondition} and the fact that $\Sigma^0_{\mu\nu}(p)$ is $O(p^4)$, whereas $\theta^0_{\mu\nu}(p)$ and $S^0_{\mu\nu}(p)$ are $O(p^2)$. The only task we are left with is to evaluate the right-hand side of \eqref{1loop at O(p^2)}, and then $\beta_1$ can be extracted from its coefficient.

\subsection{Classical flow equations}
\label{sec:Classical flow equation}

The classical flow equation \eqref{0loop} can equivalently be written as follows:
\be
    \Lambda\partial_\Lambda S^0 = \frac{1}{2\Lambda^2} \frac{\delta S^0}{\delta A_\alpha} \cdot \mathcal{K}_{\alpha\beta} \cdot \frac{\delta \Sigma^0}{\delta A_\beta} - \frac{1}{\Lambda^2} \frac{\delta S^0}{\delta A_\alpha} \cdot C_{\alpha\beta} \cdot \frac{\delta \theta^0}{\delta A_\beta} \,.
\ee
The right-hand side of the above expression has two terms with similar structure, and thus, to write the flow equation for some $n$-point vertex, one just needs to compute the contributions coming from one term as the contributions coming from the second one follow from the former by a mere relabelling. We write below the flow equations for those vertices that are relevant for the computation of $\beta_1$.

The flow equation for $S^0_{\mu_1\mu_2}(p)$ takes the following form:
\be
\label{2pt classical S0 flow equation}
    \Lambda \partial_\Lambda S^0_{\mu_1\mu_2}(p) = \frac{1}{4 \Lambda^2} S^0_{\alpha\mu_1}(p) \mathcal{K}_T(p) \Sigma^0_{\alpha\mu_2}(p) - \frac{1}{2 \Lambda^2} S^0_{\alpha\mu_1}(p) C_T(p) \theta^0_{\alpha\mu_2}(p) + (\mu_1 \leftrightarrow \mu_2) \,.
\ee
By gauge invariance and dimensions, $S^0_{\mu\nu}(p)$ (and in fact any two-point function) must have a structure similar to \eqref{S0twopt}. If we substitute this into the above flow equation and solve for some function $F(p)$, we recover the same result as stated in \eqref{S0twopt}. This is an extra check that the flow equation \eqref{2pt classical S0 flow equation} is derived in a consistent manner. 

The flow equation for $S^0_{\mu_1\mu_2\mu_3}(p_1,p_2,p_3)$ is given by:
\be
\label{3pt classical S0 flow equation}
\begin{alignedat}{2}
    \Lambda \partial_\Lambda S^0_{\mu_1\mu_2\mu_3}(p_1,p_2,p_3) &=\frac{1}{4 \Lambda^2} &\Big[ &S^0_{\alpha\mu_1}(p_1) \rK_{\mu_2;\alpha\beta}(p_2;p_1,p_3) \Sigma^0_{\beta\mu_3}(p_3) \\ 
    &&&- S^0_{\alpha\mu_1}(p_1) \rK_{\mu_3;\alpha\beta}(p_3;p_1,p_2) \Sigma^0_{\beta\mu_2}(p_2) \\
    &&&+ S^0_{\alpha\mu_1\mu_2}(p_3,p_1,p_2) \mathcal{K}_T(p_3) \Sigma^0_{\alpha\mu_3}(p_3) \\
    &&&+ S^0_{\alpha\mu_1}(p_1) \mathcal{K}_T(p_1) \Sigma^0_{\alpha\mu_2\mu_3}(p_1,p_2,p_3) \Big] \\
    &- \frac{1}{2\Lambda^2} &\Big[ &\mathcal{K}\mapsto C \text{ ; } \Sigma \mapsto \theta \Big]+ \text{cycles} \,,
\end{alignedat}
\ee
where the terms inside brackets from the last line follow from those written explicitly inside brackets above them after the relabelling indicated. One can proceed similarly to derive the flow equation for $S^0_{\mu_1\mu_2\mu_3\mu_4}(p_1,p_2,p_3,p_4)$:
\beal
\begin{split}
\label{4pt classical S0}
    \Lambda \partial_\Lambda S^0_{\mu_1\mu_2\mu_3\mu_4}(p_1,p_2,p_3,p_4) =\frac{1}{4 \Lambda^2} &\Big[ S^0_{\alpha\mu_1\mu_2\mu_3}(p_4,p_1,p_2,p_3) \mathcal{K}_T(p_4) \Sigma^0_{\alpha\mu_4}(p_4) \\
    &+ S^0_{\alpha\mu_1\mu_2}(-p_1-p_2,p_1,p_2) \mathcal{K}_{\alpha\beta}(p_1+p_2) \Sigma^0_{\beta\mu_3\mu_4}(-p_3-p_4,p_3,p_4) \\
    &+ S^0_{\alpha\mu_1}(p_1) \mathcal{K}_T(p_1) \Sigma^0_{\alpha\mu_2\mu_3\mu_4}(p_1,p_2,p_3,p_4) \\
    &+ S^0_{\alpha\mu_1}(p_1) \rK_{\mu_2;\alpha\beta}(p_2;p_1,p_3+p_4) \Sigma^0_{\beta\mu_3\mu_4}(-p_3-p_4,p_3,p_4) \\
    &- S^0_{\alpha\mu_1}(p_1) \rK_{\mu_4;\alpha\beta}(p_4;p_1,p_2+p_3) \Sigma^0_{\beta\mu_2\mu_3}(-p_2-p_3,p_2,p_3) \\
    &+ S^0_{\alpha\mu_1\mu_2}(-p_1-p_2,p_1,p_2) \rK_{\mu_3;\alpha\beta}(p_3;p_1+p_2,p_4) \Sigma^0_{\beta\mu_4}(p_4) \\
    &-  S^0_{\alpha\mu_1\mu_2}(-p_1-p_2,p_1,p_2) \rK_{\mu_4;\alpha\beta}(p_4;p_1+p_2,p_3) \Sigma^0_{\beta\mu_3}(p_3) \\
    &+ S^0_{\alpha\mu_1}(p_1) \rK_{\mu_2\mu_3;\alpha\beta}(p_2,p_3;p_1,p_4) \Sigma^0_{\beta\mu_4}(p_4) \\
    &- S^0_{\alpha\mu_1}(p_1) \rK_{\mu_2\mu_4;\alpha\beta}(p_2,p_4;p_1,p_3) \Sigma^0_{\beta\mu_3}(p_3) \\
    &- S^0_{\alpha\mu_1}(p_1) \rK_{\mu_4\mu_2;\alpha\beta}(p_4,p_2;p_1,p_3) \Sigma^0_{\beta\mu_3}(p_3) \\
    &+ S^0_{\alpha\mu_1}(p_1) \rK_{\mu_4\mu_3;\alpha\beta}(p_4,p_3;p_1,p_2) \Sigma^0_{\beta\mu_2}(p_2) \Big] \\
    - \frac{1}{2\Lambda^2} &\Big[ \mathcal{K}\mapsto C \text{ ; } \Sigma \mapsto \theta \Big] + \text{cycles} \,.
\end{split}
\eeal
In deriving these, we again exploit the interleave identities \eqref{interleave identities} up to and including the two-point level, as proven in 
 appendix \ref{appendix:interleave identities}. 
 
Although we will not need them in this paper, let us note that these equations can be integrated to give explicit formulae for the (integrated) classical effective action vertices.
Taking all the $S^{(3)}_0$ terms to the left hand side in \eqref{3pt classical S0 flow equation}, including the $S^{(3)}_0$ part of $\Sigma^{(3)}_0$, we find the same integrating factors $Z_{\mu_i \alpha}(p_i)$ as in \eqref{Z}. Thus the flow for $S^{(3)}_0$ can be similarly integrated. This results in an explicit formula for  $S^{(3)}_0$ of the same form as \eqref{sigmaint} -- except that in this case  there is also an integration constant, namely  $I_{\mu_1\mu_2\mu_3}(p_1,p_2,p_3)$. The latter follows because the $\Lambda\to\infty$ limit of $S^{(3)}_0$ is $I^{(3)}$, as \eg can be seen  from \eqref{RenormalizationCondition} and \eqref{YM}. Similarly, taking all $S^{(4)}_0$ terms to the left hand side in \eqref{4pt classical S0} gives the same integrating factors and allows its flow to be integrated up to an explicit formula for $S^{(4)}_0$. This time on the right hand side we have instances of the just computed $S^{(3)}_0$ vertex, and now the integration constant is $I^{(4)}$ as in \eqref{I4}.

\subsection{Trace terms}
\label{sec:Trace terms}

To evaluate the trace terms in \eqref{1loop at O(p^2)} above one needs to expand the log terms. The vertex expansion for the $B^\alpha{}_\beta$ terms, for example, can be done in the following way:
\beal
    \text{Tr} \ln B^\alpha{}_\beta &\equiv \text{Tr} \ln \big[ \overbrace{B_r}^{O(A_\mu^0)} + \overbrace{B_\mu(p;r,s)}^{O(A^1_\mu)} + \overbrace{B_{\mu\nu}(p,q;r,s)}^{O(A_\mu^2)} + \ldots \big] \\
    &= \int_r \ln B_r + \text{Tr} \ln \big[ 1 + B_r^{-1} B_\mu(p;r,s) + B_r^{-1} B_{\mu\nu}(p,q;r,s) + \ldots \big] \\
    &= 2N \int_q \underbrace{B_q^{-1} B_{\mu\nu}(p,-p;q,-q) - \frac{1}{2} B_q^{-1} B_\mu(p;q,-q-p) B^{-1}_{p+q} B_\nu(-p;p+q,-q)}_{O(A_\mu^2)} + \ldots \label{trace expansion B}\,,
\eeal
where $N$ accounts for group combinatorics and the extra factor of two for the $p_\mu \leftrightarrow -p_\nu$ symmetry (or, equivalently, for the definition of the two-point function). Here note that in going from the second to the third line we have discarded the log term because it is just a vacuum contribution to the flow equation (or it is differentiated away), and we have also Taylor expanded the remaining log term using 
\be \label{logexpansion}\ln (1+x) = x - \frac{x^2}{2} + \cdots\,.\ee 
Similar expressions can be written for $\big( C^{-1} \big)^a{}_b$ and $b^\alpha{}_\beta$ terms, respectively.

To evaluate the remaining trace term $\delta\Psi_0^a / \delta\phi^a$ it is useful to use its covariant representation:
\beal
    \frac{\delta \Psi_0^a}{\delta \phi^a} &= \frac{\delta}{\delta \phi^a} \bigg( -\frac{1}{2}\mathcal{K}^{ab} {\Sigma^0}_{,b} + C^{ab} {\theta^0}_{,b} \bigg) \\
    &\equiv - \frac{1}{2\Lambda^2} \frac{\delta}{\delta A_\alpha} \cdot \mathcal{K}_{\alpha\beta} \cdot \frac{\delta \Sigma^0}{\delta A_\beta} + \frac{1}{\Lambda^2} \frac{\delta}{\delta A_\alpha} \cdot C_{\alpha\beta} \cdot \frac{\delta \theta^0}{\delta A_\beta} \,.
\eeal
This allows us to compute it at the two-point level using \eqref{splitting}, yielding:
\beal
\begin{split}
\label{covrepr1}
    \frac{\delta \Psi_0^a}{\delta \phi^a} \bigg|_{\text{2-pt.function}} &= \frac{1}{2 \Lambda^2} \int_q \Bigg\{ 2 \bigg( N - \frac{1}{N} \bigg) C_{\alpha\beta}(q) \theta^0_{\beta\alpha\mu\nu}(q, -q, p, -p) \\
    &\qquad\qquad\quad - \frac{1}{N} C_{\alpha\beta}(q) \theta^0_{\beta\mu\alpha\nu}(q, p, -q, -p) \\
    &\qquad\qquad\quad + 2 N C_{\mu;\alpha\beta}(p; q, -p-q) \theta^0_{\beta\nu\alpha}(q+p, -p, -q) \\
    &\qquad\qquad\quad + 2 N C_{\mu\nu;\alpha\beta}(p, -p; q, -q) \theta^0_{\beta\alpha}(q) \\
    &\qquad\qquad\quad - 2 N C_{\mu\alpha;\alpha\beta}(p, -q; q, -p) \theta^0_{\beta\nu}(p) \Bigg\} \\
    &-\frac{1}{4\Lambda^2} \big[ C \mapsto \mathcal{K} \text{ ; } \theta \mapsto \Sigma \big] \,.
\end{split}
\eeal
This can be further refined. If we split the kernels into their transverse and longitudinal parts, respectively, then one can check explicitly that the $O(1/N)$ longitudinal pieces cancel each other after using Ward identities. One can also show that, under the $q$ integral and by using Lorentz invariance ($q \rightarrow -q$, $p_\mu \leftrightarrow -p_\nu$ \etc) together with \eqref{theta0}, \eqref{seed action}, and \eqref{4pt classical S0}, the following identities hold:
\beal
    \theta^0_{\alpha\mu\alpha\nu}(q,p,-q,-p) &=  -2\,\theta^0_{\alpha\alpha\mu\nu}(q,-q,p,-p) \,, \\  
    \Sigma^0_{\alpha\mu\alpha\nu}(q,p,-q,-p) &= -2\,\Sigma^0_{\alpha\alpha\mu\nu}(q,-q,p,-p) \,.
\eeal
This means that the $O(1/N)$ transverse pieces vanish as well, and thus \eqref{covrepr1} becomes:
\beal
\begin{split}
\label{covrepr2}
    \frac{\delta \Psi_0^a}{\delta \phi^a} \bigg|_{\text{2-pt.function}} &= \frac{N}{\Lambda^2} \int_q \Big\{ C_{\alpha\beta}(q) \theta^0_{\beta\alpha\mu\nu}(q, -q, p, -p) \\
    &\qquad\qquad + C_{\mu;\alpha\beta}(p; q, -p-q) \theta^0_{\beta\nu\alpha}(q+p, -p, -q) \\
    &\qquad\qquad + C_{\mu\nu;\alpha\beta}(p, -p; q, -q) \theta^0_{\beta\alpha}(q) \\
    &\qquad\qquad - C_{\mu\alpha;\alpha\beta}(p, -q; q, -p) \theta^0_{\beta\nu}(p) \Big\} \\
    &-\frac{N}{2\Lambda^2} \big[ C \mapsto \mathcal{K} \text{ ; } \theta \mapsto \Sigma \big] \,.
\end{split}
\eeal

\subsection{Two-point vertex at zeroth order in momentum}
\label{sec:Two-point vertex at zeroth order in momentum}

If the one-loop two-point vertex is transverse and quasi-local, as it should be, it can have no momentum independent part, \ie at  $O(p^0)$ it ought to vanish. However this property is strictly only true if the flow equation is properly regulated since, as we will see, it would then follow from the fact that this contribution can be cast as a momentum-space surface integral at large $q$, which vanishes if properly regularised.  Actually with sufficient care the construction proposed in ref. \cite{Falls:2020tmj} does ensure this, up to a divergent term that can be discarded since it is $\Lambda$ independent but differentiated by $\Lambda$, 
because it in fact depends only on two-point action vertices or zero-point kernels, as we will see. 

Thus we turn to computing the two-point $O(p^0)$ part of the one-loop flow equation \eqref{1loopB}, or equivalently \eqref{1loopA}. As we have just noted, it ought to vanish. 
The first line on the right-hand side is clearly of $O(p^2)$ and above and can therefore be discarded. This means that the only contribution comes from the functional trace terms, and hence, at $O(p^0)$, the right hand side of \eqref{1loopB} becomes:
\besp
\label{1loop at O(p^0)}
     \text{Tr} \Bigg\{ \Lambda \partial_\Lambda \Big[ \ln \big( C^{-1} \big)^a{}_b \Big] - \Lambda \partial_\Lambda \big[ \ln B^\alpha{}_\beta \big] - \frac{1}{2} \Lambda \partial_\Lambda \big[ \ln b^\alpha{}_\beta \big] + \frac{\delta \Psi_0^a}{\delta \phi^b} \\
     + \Psi_0^m \frac{\delta}{\delta \phi^m} \Big[ \ln \big( C^{-1} \big)^a{}_b \Big] - \Psi_0^m \frac{\delta}{\delta \phi^m} \big[ \ln B^\alpha{}_\beta \big] - \frac{1}{2} \Psi_0^m \frac{\delta}{\delta \phi^m} \big[ \ln b^\alpha{}_\beta \big]  \Bigg\}_{\substack{\text{2-pt.function} \\ \text{at }O(p^0)}} \,.
\eesp
Moreover, one can easily see that the second line above does not contribute at $O(p^0)$ because the blocking functional part (\ie $\Psi_0^m$) has either a two-point function residue which is at least of $O(p^2)$, or a three-point function residue which is at least $O(p)$ (furthermore it 
is multiplied by a term that is antisymmetric in $q$, and hence vanishes under the $q$ integral). The first three terms can be computed using trace expansions as described in section \ref{sec:Trace terms} above. For example, using \eqref{trace expansion B} we will obtain the following:
\be
    \text{Tr} \bigg\{ \Lambda \partial_\Lambda \big[ \ln B^\alpha{}_\beta \big] \bigg\}_{\substack{\text{2-pt.function} \\ \text{at }O(p^0)}} \equiv 2 N \Lambda \partial_\Lambda \int_q B_q^{-1} B_{\mu\nu}(0,0;q,-q) - \frac{1}{2} B_q^{-1} B_\mu(0;q,-q) B^{-1}_q B_\nu(0;q,-q) \,.
\ee
Here note that quasi-locality ensures that the $p\to0$ limit is straightforward. The above expression can be simplified by using differential Ward identities:
\beal
    \text{Tr} \bigg\{ \Lambda \partial_\Lambda \big[ \ln B^\alpha{}_\beta \big] \bigg\}_{\substack{\text{2-pt.function} \\ \text{at }O(p^0)}} &\equiv 2N \Lambda \partial_\Lambda \int_q B_q^{-1} \frac{1}{2} \partial_\mu^q \partial_\nu^q B_q - \frac{1}{2} B_q^{-1} \partial_\mu^q B_q \, B^{-1}_q \partial_\nu^q B_q \\
    &= N \Lambda \partial_\Lambda \int_q \partial_\mu^q \partial_\nu^q \big[ \ln B_q \big] \label{term1} \,.
\eeal
Similar expressions can be computed for $C$ and $b$ terms, respectively. The remaining term can be computed starting from \eqref{covrepr2} and following a similar strategy, yields:
\beal
\begin{split}
    \frac{\delta \Psi_0^a}{\delta \phi^a} \Bigg|_{\substack{\text{2-pt.function} \\ \text{at }O(p^0)}} &= \frac{N}{\Lambda^2} \int_q \Big\{ C_{\alpha\beta}(q) \frac{1}{2} \partial_\mu^q \partial_\nu^q \theta^0_{\beta\alpha}(q) \\
    &\qquad\qquad + \partial_\mu^q C_{\alpha\beta}(q) \, \partial_\nu^q \theta^0_{\beta\alpha}(q) \\
    &\qquad\qquad + \frac{1}{2} \partial_\mu^q \partial_\nu^q C_{\alpha\beta}(q) \, \theta^0_{\beta\alpha}(q) \Big\} \\
    &-\frac{N}{2\Lambda^2} \big[ C \mapsto \mathcal{K} \text{ ; } \theta \mapsto \Sigma \big] \,,
\end{split}
\eeal
which can be recast as:
\beal
    \frac{\delta \Psi_0^a}{\delta \phi^a} \Bigg|_{\substack{\text{2-pt.function} \\ \text{at }O(p^0)}} &= \frac{N}{4 \Lambda^2} \int_q \partial_\mu^q \partial_\nu^q \big[ 2 C_T(q) \theta^0_{\alpha\alpha}(q) - \mathcal{K}_T(q) \Sigma^0_{\alpha\alpha}(q) \big] \\
    &= \frac{N}{2} (D-1) \int_q \partial_\mu^q \partial_\nu^q \Big[ \tilde{q}^2 \mathcal{K}_T(q) F_{\hat{S}}(q)\Big]\\
    &= \frac{N}{2} (D-1) \int_q \Lambda \partial_\Lambda \partial_\mu^q \partial_\nu^q \big[ \ln \tilde{P}^\perp_q \big] \label{term2} \,,
\eeal
where to get the second line we used \eqref{Cthetaidentity}. The result can be cast as a total $\Lambda$-derivative which we do in the final line, where
\be 
\label{Pperp}
\tilde{P}^\perp_q := \Lambda^2 P^\perp(q)\,,\qquad P^\perp(q):= \frac1{2q^2 F(q)}\,.
\ee
As we will see in sec. \ref{sec:secondorder} this object plays the r\^ole of an effective action propagator (in the transverse space). Indeed from \eqref{S0twopt}:
\be 
\label{propagator}
P^\perp(q)\,S^0_{\alpha\beta}(q) =\delta_{\alpha\beta}-\frac{q_\alpha q_\beta}{q^2}\,.
\ee
 In \eqref{term2} one can see that the $\Lambda\partial_\Lambda$ sits inside the integral, whereas in \eqref{term1} it sits outside the momentum integral. In general, changing the integration and differentiation order is not trivial and to ensure consistency one must show that the undifferentiated expression is sufficiently well-behaved in both UV and IR. However as stressed in ref.\cite{Falls:2020tmj}, the flow equation is consistent only if we consider all trace terms together. Therefore, we should interpret this as having the $\Lambda\partial_\Lambda$ outside the integral in \eqref{term2}, as long as we add to it all the remaining terms similar to and including \eqref{term1}. This means that \eqref{1loop at O(p^0)} becomes:
\be
    \Lambda\partial_\Lambda \int_q \partial_\mu^q \partial_\nu^q \Big\{\big[ \ln C \big]_{\alpha\alpha}(q) + \ln B_q + \frac{1}{2} \ln b_q - \frac{D-1}{2} \ln \tilde{P}^\perp_q \Big\}  \,,
\ee
where $\big[ \ln C \big]_{\alpha\alpha}(q) = (D-1) \ln C_T(q) + \ln C_L(q)$. Given that the above integral is an integral of a total derivative, it amounts to a surface integral at large $q$, as advertised at the beginning of this section. This surface integral can be discarded, and thus the integral vanishes, if the term in braces above is UV finite. 
Using the definitions \eqref{C^-1}, \eqref{B} and \eqref{b}, we can write it as follows:
\beal
    (D-1) &\ln \frac{c_q}{c_q + \Tilde{q}^2} + \ln \frac{1}{1 + \Tilde{q}^2 Y_q} + \ln \Big( 1 + \Tilde{q}^2 Y_q \Big) + \frac{1}{2} \ln \Big( t_q + \tilde{q}^2 \Big) + \frac{D-1}{2} \ln \frac{2 \tilde{q}^2 \Big( c_q^2 + \tilde{q}^2 \Big)}{c_q^2} \nonumber \\
    &= - (D-1) \ln \bigg( \frac{c_q}{\tilde{q}^2} + 1 \bigg) + \frac{D-1}{2} \ln \bigg(\frac{c_q^2}{\tilde{q}^2} + 1 \bigg)  
     + \frac{1}{2} \ln \bigg( \frac{t_q}{\tilde{q}^2} + 1 \bigg) + \frac{1}{2} \ln \tilde{q}^2 \,.
\eeal
All but the last term vanish rapidly for large momentum (from \eqref{logexpansion} and because $c_q$ and $t_q$ do). In this calculation, the last term can be formally discarded because it vanishes under the combined action of $q$ and $\Lambda$ derivatives. In this sense we are justified in dropping all the $O(p^0)$ terms, confirming the arguments outlined in ref.\cite{Falls:2020tmj} work to $O(p^0)$. 

However it is unclear whether the part that leads to this unregulated $\ln\tilde{q}^2$ term can always be safely discarded. In particular it is unclear whether such a part could cause an unrecoverable failure of UV regularisation at higher loops. 
If we first compute the $q$ derivatives, the resulting momentum integral is quadratically UV divergent. By dimensions one might expect to find an analogous logarithmic UV divergence in the $O(p^2)$ part. In the next section we confirm this expectation.

\subsection{Two-point vertex at second order in momentum}
\label{sec:secondorder}

The next step is to compute the one-loop beta function by analysing the flow equation \eqref{1loop at O(p^2)}. The coefficient of the beta function at one loop is a universal quantity and thus it should be independent of all artefacts of the regularisation scheme \cite{Arnone:2002cs}. In the earlier successful construction \cite{Morris:1998kz,Morris:1999px,Morris:2000fs,Morris:2000jj,Arnone:2000bv, Arnone:2000qd,Arnone:2001iy,Arnone:2002qi,Arnone:2002fa,Arnone:2002cs,Arnone:2005fb,Arnone:2002fb,Gatti:2002kc,Morris:2005tv,Rosten:2004aw,Rosten:2005qs,Rosten:2005ka,Rosten:2006tk,Rosten:2006qx,Arnone:2006ie,Rosten:2010vm,Rosten:2011ty,Falls:2017nnu} this could be understood as follows. Since the beta function is dimensionless, the momentum integrals that compute it are also dimensionless. 
By trading the higher-point regularisation vertices in these integrals for $\Lambda$-derivatives of effective action vertices, using the classical flow equations, one finds that the results combine into terms that either vanish, because they are $\Lambda$-derivatives of regularised dimensionless integrals (which thus do not actually depend on $\Lambda$), or terms that survive but only because when cast in this way a finite result is obtained from a logarithmic IR divergence, where the effective action is universal -- as determined by the renormalization condition \eqref{RenormalizationCondition} \cite{Arnone:2002cs,Arnone:2002yh}. Here we will follow this route by manipulating the $\delta\Psi_0^a/\delta\phi^a$ term. However, before delving into the details of this, it is better for the ease of presenting to focus on the other trace terms first.

The remaining trace terms can be partitioned into two groups, namely $\Lambda\partial_\Lambda \ln K$ and $\Psi^m_0 \frac{\delta}{\delta\phi^m} \ln K$, where $K = C^{-1}, B, b$. This means that in order to compute them it will be sufficient to do this for one value of $K$ chosen for convenience, all the other contributions following from this one by substitution. It is straightforward to write down the first $B$ trace terms using \eqref{trace expansion B} as follows:
\beal
    \text{Tr} \bigg\{ \Lambda \partial_\Lambda \big[ \ln B^\alpha{}_\beta \big] \bigg\}_{\substack{\text{2-pt.function} \\ \text{at }O(p^2)}} \equiv 2 N \Lambda \partial_\Lambda \int_q &\bigg\{ B_q^{-1} B_{\mu\nu}(p,-p;q,-q) \nonumber \\
    &- \frac{1}{2} B_q^{-1} B_\mu(p;q,-q-p) B^{-1}_{p+q} B_\nu(-p;p+q,-q) \bigg\}_{O(p^2)} \label{traceterm1a}\,.
\eeal
The second $B$ trace term can be written using $\text{Tr } \Psi^m_0 \frac{\delta}{\delta\phi^m} \ln B = \text{Tr } \Psi^m_0 B_{,m} B^{-1}$, yielding:
\beal
    \text{Tr} \bigg\{ \Psi_0^m \frac{\delta}{\delta \phi^m} \big[ \ln B^\alpha{}_\beta \big] \bigg\}_{\substack{\text{2-pt.function} \\ \text{at }O(p^2)}} \equiv 2N \int_q &\bigg\{ 2 C_T(p) \theta^0_{\mu\alpha}(p) B_{\alpha\nu}(p,-p;q,-q)  B_q^{-1} \nonumber \\
    &+ C_T(p) \theta^0_{\mu\alpha}(p) B_{\alpha}(p;-p-q,q) B^{-1}_\nu(-p;p+q,-q) \nonumber \\
    &+ C_{\alpha\beta}(0) \theta^0_{\mu\nu\alpha}(p,-p,0) B_{\beta}(0;q,-q) B^{-1}_q \nonumber \\
    &- \frac{1}{2} \big[ C \mapsto \mathcal{K} \text{ ; } \theta \mapsto \Sigma \big] \bigg\}_{O(p^2)} \label{traceterm2a} \,.
\eeal
The third line is odd in $p$ and thus has no $O(p^2)$ part. Furthermore the one-point kernel is antisymmetric in $q$, whereas $B_q$ is symmetric, and so the entire row vanishes under the $q$ integral. A similar argument holds for the $\K$ sector as well. From \eqref{theta0} we get that $\theta^0_{\mu\alpha}(p) \big|_{O(p^2)} = 2\, \square_{\mu\alpha}(p)$, and, given that all the other functions appearing in \eqref{traceterm2a} are quasi-local, we are free to set $p=0$ anywhere else in the $C$ sector. In the $\K$ sector on the other hand, $\Sigma^0_{\mu\alpha}(p)$ is $O(p^4)$, which means that it does not contribute to \eqref{traceterm2a} at all. Collecting everything together we have
\beal
    \text{Tr} \bigg\{ \Psi_0^m \frac{\delta}{\delta \phi^m} \big[ \ln B^\alpha{}_\beta \big] \bigg\}_{\substack{\text{2-pt.function} \\ \text{at }O(p^2)}} \equiv 2N \int_q &\bigg\{ 2 C_T(0) 2\square_{\mu\alpha}(p) B_{\alpha\nu}(0,0;q,-q)  B_q^{-1} \label{traceterm2b} \\
    &+ C_T(0) 2\square_{\mu\alpha}(p) B_{\alpha}(0;-q,q) B^{-1}_\nu(0;q,-q) \bigg\}_{O(p^2)}  \nonumber\,,
\eeal
which we can recast if we use differential Ward identities for the kernel vertices: 
\be
\label{traceterm2c}
    \text{Tr} \bigg\{ \Psi_0^m \frac{\delta}{\delta \phi^m} \big[ \ln B^\alpha{}_\beta \big] \bigg\}_{\substack{\text{2-pt.function} \\ \text{at }O(p^2)}} \equiv 4N \square_{\mu\alpha}(p) \int_q \partial_\alpha^q \partial_\nu^q \ln B_q \,.
\ee
The flow equation \eqref{1loop at O(p^2)} shows that if we add up all trace term contributions we ought to end up with a transverse expression, \ie $\propto \square_{\mu\nu}(p)$. Terms like \eqref{traceterm2c} are transverse on the $\nu$ index if we use Lorentz invariance of the $q$ integral. But such arguments make sense strictly speaking only if the integral is properly regularised. As already noted in sec. \ref{sec:Two-point vertex at zeroth order in momentum}, trace terms should be considered together, not individually. However for more involved computations, this is not enough because the sum would still be ambiguous due to the so-called momentum routing problem (see app. \ref{app:routing}). We need in general to make statements about \emph{individual} terms, and to do so we need to apply some auxiliary regularisation for each momentum integral to give them a well defined meaning. In the end one can put all the parts back together again at which point, provided that the trace terms do actually fully regularise, the result is finite and the ``pre-regularisation'' can be safely removed. From now on we will use dimensional regularisation for this purpose, \ie compute in $d=4-\varepsilon$ dimensions.  This means that terms similar to \eqref{traceterm2c} are indeed transverse on both indices. If we contract $p_\mu$ into \eqref{traceterm1a} and use Ward identities we will get (ignore the $O(p^2)$ requirement for now):
\beal
\label{relabelexample}
    &2 N \Lambda \partial_\Lambda \int_q \bigg\{ B_q^{-1} \Big( B_\mu(0;q,-q) - B_\mu(-p;p+q,-q) \Big) \nonumber \\
    &\qquad\qquad\quad - \frac{1}{2} B_q^{-1} \big( B_q - B_{p+q} \big) B^{-1}_{p+q} B_\nu(-p;p+q,-q) \bigg\} \,.
\eeal
The first term in the bracket on the first line is odd in $q$, and thus vanishes when integrated. In the second line we can relabel $q \rightarrow -q-p$ for the first term inside the bracket and use the antisymmetry of the one-point kernel vertex, $B_\nu(-p;p+q,-q) = - B_\nu(-p;-q,p+q)$, to arrive at
\beal
    &2 N \Lambda \partial_\Lambda \int_q \bigg\{ - B_q^{-1} B_\mu(-p;p+q,-q) + B_q^{-1} B_\nu(-p;p+q,-q) \bigg\} \label{tracetermbefore1b} \\
    =& 2 N \Lambda \partial_\Lambda \int_q \big\{ 0 \big\} \label{traceterm1b}\,.
\eeal
Lorentz invariance implies that a similar result would have been obtained if $p_\nu$ had been used. Therefore, all terms with a similar structure to \eqref{traceterm1a}, when properly regularised, are individually transverse.  

The last step is to compute $\delta\Psi_0^a/\delta\phi^a$. At this point we note that \eqref{covrepr2} is formally transverse prior to any substitution being done. One can easily check this by contracting it with $p_\mu$ (or equivalently $p_\nu$), and then show that the expression vanishes at all orders, not only at $O(p^2)$. This means that formally all trace terms in \eqref{1loop at O(p^2)} are individually transverse. This is gratifying but unsurprising: it is a consistency check on the formal preservation of gauge invariance by the flow equation. As we have just emphasised, for these manipulations to be meaningful we need to verify this with a gauge invariant pre-regularisation in place. If the expressions are in fact properly regularised through the PV trace terms, we should then be able to combine the results into a finite transverse, and in fact universal, answer.

Thus the real test is to see whether \eqref{covrepr2} remains transverse in dimensional regularisation after we trade four and three-point $\Sigma^0$ and $\theta^0$ vertices for ($\Lambda$ derivatives of) effective action vertices. If we substitute first \eqref{4pt classical S0} into \eqref{covrepr2} to trade four-point $\Sigma^0$ and $\theta^0$ vertices, and then use \eqref{3pt classical S0 flow equation} to trade away the $\Sigma^0$ and $\theta^0$ three-point vertices, 
we obtain eventually the following:
\beal
\begin{split}
\label{traceterm3}
    \frac{\delta \Psi_0^a}{\delta \phi^a} \bigg|_{\substack{\text{2-pt.function} \\ \text{at }O(p^2)}} = N \int_q \Bigg\{ &-\Lambda\partial_\Lambda \Big[ P^\perp(q) S^0_{\alpha\alpha\mu\nu}(q,-q,p,-p) \Big]_{O(p^2)} \\
    &+ \frac{1}{2} \Lambda\partial_\Lambda \Big[ P^\perp(q) P^\perp(p-q) S^0_{\alpha\beta\mu}(p-q,q,-p) S^0_{\alpha\beta\nu}(p-q,q,-p) \Big]_{O(p^2)} \\
    &+ \frac{\square_{\alpha\nu}(p)}{\Lambda^2 q^2 F_q} \Bigg[ \frac{\big( 1-\mathcal{K}_T(q) \big)\Sigma^0_{\alpha\mu}(q)-\big( 1-2C_T(q) \big)\theta^0_{\alpha\mu}(q)}{q^2} \\
    &\qquad\qquad\quad - \frac{1}{2} \partial_\mu^q\partial_\alpha^q \big( S^0_{\beta\beta}(q) - \Sigma^0_{\beta\beta}(q) + \theta^0_{\beta\beta}(q) \big) \\
    &\qquad\qquad\quad + 2\frac{q_\alpha q_\mu}{q^2} \big( \mathcal{K}_T(q) F_\Sigma(q) - 2 C_T(q) F_\theta(q) - F_q \big) \\
    &\qquad\qquad\quad + \frac{1}{2 \Lambda^2 q^2 F_q} \partial_\mu^q \big(\square_{\beta\gamma}(q) F_q \big) \, \partial_\alpha^q \big( S^0_{\beta\gamma}(q) - \Sigma^0_{\beta\gamma}(q) + \theta^0_{\beta\gamma}(q) \big) \Bigg] \Bigg\} \,,
\end{split}
\eeal
where $P^\perp(q)$ is an effective propagator and was defined already in \eqref{Pperp}. To arrive at this expression we use the symmetries including differential Ward identities in a similar way to the trace terms above. As there, we do not need the explicit formulae except for the zero-point kernels and two-point action vertices.

All the terms appearing in the above expression are manifestly transverse except the $\Lambda$-derivative ones. We can evaluate the contribution of the latter in dimensional regularisation and check whether they are transverse or not. Again, we are free to move the $\Lambda$ derivative outside the integral if the resulting momentum integrals are regularised. The first term (containing the four-point vertex) would then vanish if the UV regularisation (as also provided by the trace terms) is correctly in place,  because the integral is dimensionless in four dimensions and thus actually independent of $\Lambda$. This is so because, although $P^\perp\sim1/q^2$ for small $q$, this IR divergence is integrable in four dimensions. Similarly the second term (containing the three-point effective action vertices) has no surviving UV contribution if it is properly regulated there, but this time there is a logarithmic IR divergence which when differentiated with respect to $\Lambda$ gives a finite universal answer dependent only on the renormalization condition \eqref{RenormalizationCondition}.

Now we contract the first two terms in \eqref{traceterm3} with $p_\mu$ and use Ward identities to further simplify it, following essentially the same steps that took us from \eqref{relabelexample} to \eqref{tracetermbefore1b}. This time we get:
\be \label{longitudinalbefore}
 N\Lambda\partial_\Lambda \int_q P^\perp(q) \Bigg\{ S^0_{\alpha\alpha\nu}(q,p-q,-p) - P^\perp(p-q) S^0_{\alpha\beta}(p-q) S^0_{\alpha\beta\nu}(q,p-q,-p) \Bigg\}\,.
\ee
Now using the propagator identity \eqref{propagator} and Ward identities recursively this simplifies to
\be
    \Lambda\partial_\Lambda \int_q \frac{(p+q)_\beta}{(p+q)^2} \bigg( \delta_{\beta\nu} - \frac{q_\beta q_\nu}{q^2} \bigg) \,.
\ee
The $\delta_{\beta\nu}$ part is odd in $q$ (one can easily see that this is the case after expanding the bracket and relabelling $q \rightarrow q-p$), and hence vanishes, whereas the second term can be recast using $q\!\cdot\!(p+q) = \frac{1}{2}[(p+q)^2+q^2-p^2]$, yielding:
\be
\label{unregulated}
    -\frac{\Lambda}2\frac{\partial}{\partial\Lambda} \int_q \frac{q_\nu}{q^2} + \frac{q_\nu}{(p+q)^2} - \frac{p^2 q_\nu}{(p+q)^2 q^2} \,.
\ee
Again, the first term is odd in $q$. The second term is quadratically divergent but in dimensional regularisation it vanishes, whereas the third term is linearly divergent and gives a finite contribution in dimensionsal regularisation (due to a logarithmic subdivergence):
\be
    \Lambda\partial_\Lambda \bigg( - \frac{1}{2} \frac{p^2 p_\nu}{(4\pi)^2} \ln \Lambda \bigg) = -\frac{1}{2} \frac{p^2 p_\nu}{(4\pi)^2} \,,
\ee
where we recognise that the regularisation scale is set by $\Lambda$.
This means that the original $\Lambda$ derivative terms from \eqref{traceterm3} give a non-vanishing longitudinal contribution in dimensional regularisation, which amounts to
\be
\label{longitudinalpart}
    -\frac{1}{2} \frac{p_\mu p_\nu}{(4\pi)^2} \,,
\ee
and thus they cannot be transverse on their own. In fact, apart from an overall factor of $\frac{1}{2}$, the effective action terms in \eqref{traceterm3} have precisely the same structure as found in ref.\cite{Arnone:2002cs} as part of the one-loop beta function computation using the manifestly gauge invariant flow equation developed there. Moreover, the computed longitudinal part \eqref{longitudinalpart} is precisely half the one found in ref. \cite{Arnone:2002cs}. The same arguments in that paper can be used to extract the full IR contribution from the first two lines of \eqref{traceterm3}:
\be 
\frac{19}{6}p^2\delta_{\mu\nu} -\frac{11}3p_\mu p_\nu\,,
\ee
and this has the same longitudinal part.

Thus we see that if we assume that \eqref{1loop at O(p^2)} is sufficiently regularised in the UV we get a non-transverse answer.  As we noted, the above result is arrived at independently of the detailed form of the regularisation structure, apart from the expressions for two-point action vertices and zero-point kernels. This means that the problem cannot be cured by a more careful choice of cutoff functions or an alternative covariantization, but lies at a deeper structural level.

Since formally the integrals start out transverse, it follows that there are unregulated UV divergences which provide a cancelling longitudinal contribution. In fact if the third term in \eqref{unregulated} has no UV regularisation we can (cavalierly) regard it as an infinite constant, independent of $\Lambda$, which is then annihilated by the $\Lambda$ derivative. 

The problem then is that we cannot extract universal information from the $S_0$ terms in \eqref{traceterm3}. We can see this explicitly by substituting $S_0=\Sigma_0+\hat{S}$. Since $P^\perp(q) \approx c^2_q$, \cf \eqref{Pperp} and \eqref{S0twopt}, the resulting terms with a $\Sigma_0$ three-point or four-point vertex are UV regularised because $\Sigma^{(3)}_0$ and $\Sigma^{(4)}_0$  diverge only as $c^{-1}_q$, as we confirmed explicitly in sec. \ref{sec:Regularisation of higher point classical vertices}.  As explained above this means that the $\Sigma^{(4)}_0$ contribution actually vanishes, because the integral is then well defined and dimensionless, and thus annihilated by $\Lambda\partial_\Lambda$. The contributions with a $\Sigma^{(3)}_0$ vertex vanish for the same reason. To see this we only need to show that there is no longer a logarithmic divergence at small $q$. Since $\Sigma^{(2)}_0$ is $O(p^4)$, \cf \eqref{Sigma2pt}, we know by gauge invariance (or equivalently the Ward identities) that
 $\Sigma^0_{\alpha\beta\mu}(p-q,q,-p)$ starts cubic in momenta in a small momentum expansion. In other words, retaining powers of $p$ up to a maximum of $O(p^2)$, as $q\to0$ it vanishes with at least one power of $q$, and thus indeed  there is no longer an IR logarithmic divergence from the second line of \eqref{traceterm3}.
 
This means that, on substituting $S_0=\Sigma_0+\hat{S}$, the top two lines of \eqref{traceterm3} become precisely the same expression but with $S^{(3)}_0$ and $S^{(4)}_0$ replaced by $\hat{S}^{(3)}_0$ and $\hat{S}^{(4)}_0$ respectively. However $\hat{S}$ is freely designed by us as part of the regularisation scheme so cannot of itself contain the requisite universal information. Indeed we can go one step further and introduce a transverse space effective propagator for $\hat{S}_{\alpha\beta}(q)$:
\be \hat{P}(q) := \frac1{2q^2F_{\hat{S}}(q)}\,, \ee
which satisfies the obvious transverse projector relation, \cf \eqref{Shattwopt} and  \eqref{propagator}. Substituting 
\be 
P^\perp(q) = \hat{P}(q)\, +\left[ P^\perp(q) - \hat{P}(q) \right] = \hat{P}(q)\, + \frac{c^3_q}{2q^2}\frac{1-c_q}{(c^2_q+\tilde{q}^2)(c_q+\tilde{q}^2)}\,,
\ee
for all propagators in the same top two lines, it is apparent that any part containing the last expression above is again both UV and IR finite (in the latter case because $c(0)=1$) and therefore vanishes under the $\Lambda$ derivative. Thus the top two lines only contain information determined by our choice of seed action.

We have shown that if we assume that \eqref{1loop at O(p^2)} is sufficiently regularised in the UV we get a universal non-transverse answer from a certain IR contribution. Since formally the integrals start out transverse, it follows that there are actually unregulated UV divergences which can formally be viewed as providing a cancelling longitudinal contribution. Although one can indeed proceed in this way, it is then no longer possible to prove the result is universal. Indeed we show that this contribution can be expressed solely in terms of the seed action, a quantity that we are free to choose.

\section{Summary and Conclusions}
\label{sec:Summary and Conclusions}

We finish this paper with a brief summary and discussion of our main findings. 
In sec. \ref{sec:Two-point vertex at zeroth order in momentum}, we showed that all but one term in the $O(p^0)$ part of the two-point function at one-loop level is correctly regularised. Key to this, was the finding that the $O(p^0)$ actually only depends on zero-point kernels and two-point action vertices. If properly regularised, it has to vanish by gauge invariance. We show that it does vanish. For the properly regularised terms, this is so because they can be written as a vanishing UV surface term in momentum space. The one term that is not properly regularised can however be formally set to zero because it is the $\Lambda$-derivative of a $\Lambda$-independent quadratically divergent term.

The $O(p^2)$ part should give us the one-loop beta function, if properly regularised, and thus be universal (independent of regularisation artefacts).   
In sec. \ref{sec:secondorder} we recast the result in a way that has previously allowed a universal answer to be extracted, if properly regularised \cite{Arnone:2002cs,Arnone:2002yh}. We saw that although the $O(p^2)$ part is formally transverse, if computed assuming complete UV regularisation there is actually a non-vanishing longitudinal part. The problem is caused by unregulated UV divergences, which appears to be associated to the one found at $O(p^0)$, and is independent of choice of cutoff functions or covariantization. 

Similarly to the $O(p^0)$ case, if the incompletely regularised longitudinal part is regarded formally as an infinite constant, it can then be set to zero, since it is differentiated by $\Lambda$. The problem then is that  the earlier techniques can no longer be used to extract a universal answer because it is no longer clear which IR divergent terms should provide a universal answer and which should be formally set to zero. Indeed we saw explicitly that in the current case these terms can then be made independent of the classical effective action. Instead they can be seen to depend only on the seed action, part of the regularisation structure which we are free to choose (and which is thus non-universal).

Note that these results are not  at variance with ref. \cite{Falls:2020tmj} where the one-loop beta function coefficient is successfully computed. It is there extracted from determinants of zero-point kernels and two-point vertices (the result obtained in \ref{sec:Two-point vertex at zeroth order in momentum}), using a heat kernel approach. It is thus insensitive to the problems with UV regularisation that we are highlighting. Indeed the heat-kernel expression formally depends only on the value of these kernels at vanishing momentum.

Our results show that the one-loop off-shell contribution is not fully regularised. At one loop it appears to be possible to proceed formally and obtain valid answers, but it is unclear whether that would still be possible at higher loops. Also, the powerful techniques \cite{Arnone:2002cs,Arnone:2002yh} that previously allowed universal results to be extracted,  unfortunately fail in this formulation. As we have already noted, a repair of this problem (if there is one which does not introduce explicit PV fields as done previously \cite{Morris:1998kz,Morris:1999px,Morris:2000fs,Morris:2000jj,Arnone:2000bv, Arnone:2000qd,Arnone:2001iy,Arnone:2002qi,Arnone:2002fa,Arnone:2002cs,Arnone:2005fb,Arnone:2002fb,Gatti:2002kc,Morris:2005tv,Rosten:2004aw,Rosten:2005qs,Rosten:2005ka,Rosten:2006tk,Rosten:2006qx,Arnone:2006ie,Rosten:2010vm,Rosten:2011ty,Falls:2017nnu}) would seem to require structural changes. Hints lie in the insufficient regularisation at $O(p^0)$ seen in \ref{sec:Two-point vertex at zeroth order in momentum}, and in the longitudinal sector as seen in sec. \ref{sec:secondorder} and discussed in more general terms in sec. \ref{sec: regularisation failure sketch}.  
Hopefully, following these hints, a way can be found to implement a more complete  regularisation.

\vskip2cm
\section*{Acknowledgments}

We thank Kevin Falls for useful comments on an earlier version of this manuscript.
VMM acknowledges support via an STFC PhD studentship. TRM acknowledges support from STFC through Consolidated Grant ST/T000775/1.


\appendix

\section{Interleave identities for compound kernels}
\label{appendix:interleave identities}

Interleave identites \eqref{interleave identities} hold for those kernels whose covariantization is carried by gauge fields acting by commutation. This means, for example, that any well-behaved (or quasi-local) function of $\Delta^a{}_b$ or $\Delta_\parallel$ fulfills this condition. We exploit these identities for example in the one-loop beta function calculation where it allows the cancellation of certain terms.
However, the two kernels $\K^a{}_b$ and $C^a{}_b$, are not solely constructed in this way, and it is not a priori evident that they still obey interleave identities. It may be that they do not at some higher point level. Fortunately for us they do obey the interleave identities up to the two-point level, and this is all that we need in this paper. 

In this appendix we prove that the identities do hold up to this level using the other already established symmetries. At the same time we give more details on how to construct them. Given the definition of $\K^a{}_b$ in \eqref{Kkernel}, it is sufficient to check the vertices of $C^a{}_b$. Then, since $\kappa^a{}_b$ is a function of $\Delta^a{}_b$, it is guaranteed that a similar behaviour is inherited by the $\K^a{}_b$ vertices.  
From \eqref{C^-1} we can write $ \big( C^{-1} \big)^a{}_b$ in the following way:
\be
\label{newC^-1}
    \big( C^{-1} \big)^a{}_b = \delta^a{}_b + R^a{}_b + z^a{}_b \,,
\ee
where $R^{ab} := \gamma^{am} A_{,mn} \gamma^{nb}$ and $z^{ab} := K^a_\alpha Y^{\alpha\beta} K^b_\beta$. It is easy to see that $z^a{}_b$ is made out of gauge fields which act by commutation, and hence its vertices obey interleave identities. However, checking whether $R$ vertices satisfy these identities or not is non-trivial. 
From sec. \ref{sec:Kernel vertices} we see that
for two (matrix valued) functions $J_\mu(x)$ and $J_\nu(y)$ the following equivalent statements can be written:
\beal
    J_a R^{ab} J_b &\equiv \int_x \int_y J^a_\mu(x) R^{ab}_{\mu\nu}(x,y) J^b_\nu(y) \label{representation1} \\
    &= \frac{2}{\Lambda^2} \text{tr} \int_x \int_y J_\mu(x) R_{\mu\nu}(x,y) J_\nu(y) \label{representation2a} \\
    &= \frac{2}{\Lambda^2} \sum_{k=0}^\infty \sum_{m=0}^k \int_x \int_y \int_{x_1 \cdots x_k} R_{\mu_1 \ldots \mu_{k-m} , \mu_{k-m+1} \ldots \mu_k ;\mu\nu}(x_1 \ldots x_{k-m}; x_{k-m+1} \ldots x_k ; x, y) \nonumber \\
    &\hspace{1.5cm} \times \text{tr} \Big[ J_\mu(x) A_{\mu_1}(x_1) \cdots A_{\mu_{k-m}}(x_{k-m}) J_\nu(y) A_{\mu_{k-m+1}}(x_{k-m+1}) \cdots A_{\mu_k}(x_k) \Big] \label{representation2b} \,,
\eeal
where we note that the sum is half that in \eqref{KernelExplicitRep} and we used the definition of functional derivatives below \eqref{functional} and  normalisation of the generators. We also note that
$R^{ab}_{\mu\nu}(x,y)$ is given by\footnote{Recall that we factor out the powers of $g$, \cf sec. \ref{sec:Loop expansion}.}
\be
\label{explicit T}
    R^{ab}_{\mu\nu}(x,y) =\frac1{\Lambda^4} \frac{\delta^2 A}{\delta A^b_\nu(y)\delta A^a_\mu(x)} \,,
\ee
We can look at the functional derivatives of $A$ in more detail if we use the trace expansion \eqref{S trace expansion}. The first functional derivative takes the following form:
\be
    \frac{\delta A}{\delta A^a_\mu(x)} = \sum_{k=2}^\infty \int_{x_1 \ldots x_{k-1}} A_{\mu \mu_1 \ldots \mu_{k-1}}(x, x_1 \ldots x_{k-1}) \text{tr} \Big[ T^a A_{\mu_1}(x_1) \ldots A_{\mu_{k-1}}(x_{k-1}) \Big] \,,
\ee
where we consider only the single-trace terms which are the only ones of interest to us. If we differentiate it a second time, we will obtain:
\besp
    \frac{\delta^2 A}{\delta A^b_\nu(y)\delta A^a_\mu(x)}= \\ 
    =\sum_{k=2}^\infty \sum_{m=0}^{k-2} \int_{x_1 \ldots x_{k-2}} A_{\mu \mu_1 \ldots \mu_{k-2-m} \nu \mu_{k-1-m} \ldots \mu_{k-2}}(x, x_1 \ldots x_{k-2-m}, y, x_{k-1-m} \ldots x_{k-2}) \\
    \text{tr} \Big[ T^a A_{\mu_1}(x_1) \ldots A_{\mu_{k-2-m}}(x_{k-2-m}) T^b A_{\mu_{k-1-m}}(x_{k-1-m}) \ldots A_{\mu_{k-2}}(x_{k-2}) \Big] \,.
\eesp
If we now relabel $k \mapsto k+2$ and substitute the result into \eqref{representation1} using \eqref{explicit T}, we will obtain the following identity:
\besp
    J_a R^{ab} J_b \equiv \frac{1}{\Lambda^4} \sum_{k=0}^\infty \sum_{m=0}^k  \int_x \int_y \int_{x_1 \ldots x_k} A_{\mu \mu_1 \ldots \mu_{k-m} \nu \mu_{k-m+1} \ldots \mu_k}(x, x_1 \ldots x_{k-m}, y, x_{k-m+1} \ldots x_k) \\
    \text{tr} \Big[ J_\mu(x) A_{\mu_1}(x_1) \ldots A_{\mu_{k-2-m}}(x_{k-m}) J_\nu(y) A_{\mu_{k-m+1}}(x_{k-m+1}) \ldots A_{\mu_k}(x_k) \Big] \,.
\eesp
This means that \eqref{representation1} and \eqref{representation2b} remain equivalent provided that we make the following identification:
\besp
    R_{\mu_1 \ldots \mu_{k-m} , \mu_{k-m+1} \ldots \mu_k ;\mu\nu}(x_1 \ldots x_{k-m}; x_{k-m+1} \ldots x_k ; x, y) = \\
    = \frac1{2\Lambda^2} A_{\mu \mu_1 \ldots \mu_{k-m} \nu \mu_{k-m+1} \ldots \mu_k}(x, x_1 \ldots x_{k-m}, y, x_{k-m+1} \ldots x_k) \,,
\eesp
which take the exact same form in momentum space.

If we now look at the one-point $R$ part of \eqref{newC^-1}, we will obtain:
\beal
    R_{\mu;\alpha\beta}(p;r,s) &= \frac1{2\Lambda^2} A_{\mu\beta\alpha}(p,s,r) \,, \\
    R_{,\mu;\alpha\beta}(;p;r,s) &= \frac1{2\Lambda^2} A_{\mu\alpha\beta}(p,r,s) \,. 
\eeal
Three-point action vertices are totally anti-symmetric in all their arguments, and thus the above expressions can be recast as
\be
\label{id1}
    R_{\mu;\alpha\beta}(p;r,s) + R_{,\mu;\alpha\beta}(;p;r,s) = 0 \,,
\ee
which is exactly \eqref{interleave identities} at the one-point level. 

At the two-point level the $R$ part of \eqref{newC^-1} becomes: 
\beal
    R_{\mu\nu;\alpha\beta}(p,q;r,s) &= \frac1{2\Lambda^2} A_{\mu\nu\beta\alpha}(p,q,s,r) \,, \label{R2a}\\
    R_{\mu,\nu;\alpha\beta}(p;q;r,s) &= \frac1{2\Lambda^2} A_{\mu\beta\nu\alpha}(p,s,q,r) \,, \label{R2b}\\
    R_{,\mu\nu;\alpha\beta}(;p,q;r,s) &= \frac1{2\Lambda^2} A_{\mu\nu\alpha\beta}(p,q,r,s) \,. \label{R2c} 
\eeal
From \eqref{R2a} and \eqref{R2c} we can write the following:
\beal
\begin{split}
\label{id2}
    R_{,\mu\nu;\alpha\beta}(;p,q;r,s) - R_{\nu\mu;\alpha\beta}(q,p;r,s) &= \frac1{2\Lambda^2} A_{\mu\nu\alpha\beta}(p,q,r,s) - \frac1{2\Lambda^2} A_{\nu\mu\beta\alpha}(q,p,s,r) \\
    &= \frac1{2\Lambda^2} A_{\mu\nu\alpha\beta}(p,q,r,s) - \frac1{2\Lambda^2} A_{\mu\nu\alpha\beta}(p,q,r,s) = 0 \,,
\end{split}
\eeal
where in going from the first to the second line we have used charge conjugation invariance. Similarly, working out the explicit form of the four-point $A$ vertices, one can show that the following expression also holds:
\be
\label{id3}
    R_{\mu,\nu;\alpha\beta}(p;q;r,s) + R_{\mu\nu;\alpha\beta}(p,q;r,s) + R_{\nu\mu;\alpha\beta}(q,p;r,s) = 0 \,.
\ee
These last two identities are exactly the interleave identities \eqref{interleave identities} at the two-point level for the $R$ part of $C^{-1}$. This means that the full one-point and two-point vertices of $C^{-1}$ also obey interleave identities.

We are now at the stage where we can inspect the vertices of $C^a{}_b$ using $C^a{}_m \big( C^{-1} \big)^m{}_b = \delta^a{}_b$. At $O(A_\mu)$, this can be expressed schematically in the following way:
\be
    \text{one-point } C = - \Big( \text{zero-point } C \Big) \times \bigg( \text{one-point } C^{-1} \bigg) \times \Big( \text{zero-point } C \Big) \,, 
\ee
or explicitly as a set of two identities:
\beal
    C_{\mu;\alpha\beta}(p;r,s) &= - C_{\alpha\gamma}(r) C^{-1}_{\mu;\gamma\delta}(p;r,s) C_{\delta\beta}(s) \,, \\ C_{,\mu;\alpha\beta}(p;r,s) &= - C_{\alpha\gamma}(r) C^{-1}_{,\mu;\gamma\delta}(p;r,s) C_{\delta\beta}(s) \,.
\eeal
If we use \eqref{id1} for the full one-point $C^{-1}$ vertices, we can recast the expressions above as:
\be
    C_{\mu;\alpha\beta}(p;r,s) + C_{,\mu;\alpha\beta}(;p;r,s) = 0 \,,
\ee
which is exactly the interleave identity \eqref{interleave identities} for one-point $C$ vertices. 

Similarly, at $O(A^2_\mu)$ one can write schematically:
\besp
    \text{two-point } C =  - \Big( \text{zero-point } C \Big) \times \bigg( \text{two-point } C^{-1} \bigg) \times \Big( \text{zero-point } C \Big) \\
    + \Big( \text{zero-point } C \Big) \times \bigg( \text{one-point } C^{-1} \bigg) \times \Big( \text{zero-point } C \Big) \\
    \times \bigg( \text{one-point } C^{-1} \bigg) \times \Big( \text{zero-point } C \Big) \,.
\eesp
 From the above expression we can infer the explicit expression for the two-point $C$ vertices:
\beal
\begin{split}
\label{expr11}
    C_{\mu,\nu;\alpha\beta}(p;q;r,s) = &- C_{\alpha\alpha_1}(r) C^{-1}_{\mu,\nu;\alpha_1\beta_1}(p;q;r,s) C_{\beta_1\beta}(s) \\
    &+ C_{\alpha\alpha_1}(r) C^{-1}_{\mu;\alpha_1\gamma}(p;r,s+q) C_{\gamma\delta}(s+q) C^{-1}_{,\nu;\delta\beta_1}(;q;p+r,s) C_{\beta_1\beta}(s) \\
    &+ C_{\alpha\alpha_1}(r) C^{-1}_{,\nu;\alpha_1\gamma}(;q;r,s+p) C_{\gamma\delta}(r+q) C^{-1}_{\mu;\delta\beta_1}(p;q+r,s) C_{\beta_1\beta}(s) \,,
\end{split}
\\
\begin{split}
\label{expr20}
    C_{\mu\nu;\alpha\beta}(p,q;r,s) = &- C_{\alpha\alpha_1}(r) C^{-1}_{\mu\nu;\alpha_1\beta_1}(p,q;r,s) C_{\beta_1\beta}(s) \\
    &+ C_{\alpha\alpha_1}(r) C^{-1}_{\mu;\alpha_1\gamma}(p;r,s+q) C_{\gamma\delta}(p+r) C^{-1}_{\nu;\delta\beta_1}(q;p+r,s) C_{\beta_1\beta}(s) \,,
\end{split}
\\
\begin{split}
\label{expr02}
    C_{,\mu\nu;\alpha\beta}(;p,q;r,s) = &- C_{\alpha\alpha_1}(r) C^{-1}_{,\mu\nu;\alpha_1\beta_1}(;p,q;r,s) C_{\beta_1\beta}(s) \\
    &+ C_{\alpha\alpha_1}(r) C^{-1}_{,\mu;\alpha_1\gamma}(;p;r,s+q) C_{\gamma\delta}(p+r) C^{-1}_{,\nu;\delta\beta_1}(;q;p+r,s) C_{\beta_1\beta}(s) \,.
\end{split}
\eeal
The interleave identities $C_{,\mu\nu;\alpha\beta}(;p,q;r,s) - C_{\nu\mu;\alpha\beta}(q,p;r,s) = 0$ and $C_{\mu,\nu;\alpha\beta}(p;q;r,s) + C_{\mu\nu;\alpha\beta}(p,q;r,s) + C_{\nu\mu;\alpha\beta}(q,p;r,s) = 0$ follow trivially from \eqref{expr11}, \eqref{expr20}, and \eqref{expr02} if we use \eqref{id1}, \eqref{id2} for the full $C^{-1}$ vertices. This concludes the proof that the one and two-point functions of $C^a{}_b$ and $\K^a{}_b$, respectively, obey interleave identities.

\section{Why preregularisation is necessary}
\label{app:routing}

Consider the momentum integral \cite{Morris:1998kz} 
\be\label{vanish} \int_q \left\{ \frac1{q^2+\Lambda^2}-\frac1{(q+p)^2+\Lambda^2} \right\}\,. \ee
Evidently this is zero, since in the second term we can change variables to $q\mapsto-q-p$ to make it equal and opposite to the first. These sort of cancelling terms are generic and unavoidable in a system of regularisation that uses gauge invariant PV regularisation. For example, just such a relabelling is used in \eqref{relabelexample} to cancel terms against each other. Depending on how terms are unpacked and combined, they can then give contributions that appear in a similar way to above. Expanding the second term in \eqref{vanish} to $O(p^2)$, we obtain
\be \label{vanishp2} \int_q\left\{ \frac{p^2}{(q^2+\Lambda^2)^2} -4\frac{(p\!\cdot\!q)^2}{(q^2+\Lambda^2)^3}\right\}\,.\ee
In four dimensions by Lorentz invariance (actually rotational invariance since we are in Euclidean signature) we can replace $(p\!\cdot\!q)^2$ by $q^2p^2/4$, so the above result simplifies to
\be \label{expanded} \Lambda^2p^2\int_q \frac1{(q^2+\Lambda^2)^3}\,,\ee
a manifestly positive convergent answer from a vanishing integral! The problem is that \eqref{vanish} is finite but ambiguous; the result \eqref{expanded} can be cast as a total derivative but with a finite surface term. Generally in situations where PV regularisation is used, we must first pre-regularise (in a way that is compatible with gauge invariance) to avoid this so-called momentum routing problem. This is done, not to subtract divergences with respect to the pre-regulator, but in order to ensure that the integrals are defined sufficiently carefully \cite{Bakeyev:1996is,Asorey:1995tq,Warr:1986we}. In this paper we use dimensional regularisation, working in $d=4-\varepsilon$ dimensions when necessary.
Then in \eqref{vanishp2}, we effectively replace $(p\!\cdot\!q)^2$ by $q^2p^2/4 +\varepsilon\, q^2p^2/16$. Using dimensional regularisation it is straightforward to evaluate the new contribution and verify that it cancels \eqref{expanded}, restoring agreement with \eqref{vanish}.


\bibliographystyle{hunsrt}
\bibliography{references}
\end{document}